\documentclass[12pt]{iopart}
\usepackage{iopams}
\usepackage{amsfonts}
\usepackage{graphicx}
\usepackage{float}
\usepackage{cite}

\newcommand{\bt}[1]{\mathbf{#1}}
\newcommand{\bo}[1]{\boldsymbol{#1}}

\begin{document}
	
\title[]{Modified transverse Ising model for the dielectric properties of SrTiO$_3$ films and interfaces}
\author{Kelsey S Chapman$^1$ and W A Atkinson$^1$} 
\address{$^1$Department of Physics and Astronomy, Trent University, Peterborough, Ontario, Canada, K9L 0G2}
\ead{kelseychapman@trentu.ca, billatkinson@trentu.ca}
\date{\today}

\begin{abstract}
The transverse Ising model (TIM), with pseudospins representing the lattice polarization, is often used as a simple description of ferroelectric materials. However, we demonstrate that the TIM, as it is usually formulated, provides an incorrect description of SrTiO$_{3}$ films and interfaces because of its inadequate treatment of spatial inhomogeneity. We correct this deficiency by adding a pseudospin anisotropy to the model. We demonstrate the physical need for this term by comparison of the TIM to a typical Landau-Ginzburg-Devonshire model. We then demonstrate the physical consequences of the modification for two model systems: a ferroelectric thin film, and a metallic LaAlO$_{3}$/SrTiO$_{3}$ interface. We show that, in both cases, the modified TIM has a substantially different polarization profile than the conventional TIM. In particular, at low temperatures the formation of quantized states at LaAlO$_{3}$/SrTiO$_{3}$ interfaces only occurs in the modified TIM.
\end{abstract}

\vspace{2pc}

\submitto{\JPCM}

\maketitle
%\ioptwocol

\section{Introduction}

The transverse Ising model (TIM) was developed by deGennes in 1963 to describe the ferroelectric transition in hydrogen-bonded materials like potassium dihydrogen phosphate (KDP) \cite{degennes63}. As suggested by its name, the model formally describes a system of magnetic Ising moments in a transverse magnetic field \cite{stinchcombe73},
and since its discovery it has become significant because it is one of the simplest models to exhibit a quantum phase transition \cite{Sachdev:2011}. The focus of this work is more practical; we explore the use of the TIM to describe the dielectric properties of SrTiO$_3$. Indeed, the TIM has been used widely to model the low-energy physics of systems in which local degrees of freedom can be represented by pseudospins \cite{stinchcombe73}. In KDP, for example, the $S=\frac{1}{2}$ Ising spin states represent the two degenerate positions available to each hydrogen atom, while the transverse field represents the quantum mechanical tunneling between the states.

Because the TIM starts from a picture of fluctuating local dipole moments, it naturally describes materials, like KDP, with order-disorder transitions.  However, the model has also been applied to materials like SrTiO$_3$, which are close to a displacive ferroelectric transition.  While there are some clear discrepancies between the model and experiments \cite{Muller:1979wa}, the mean-field TIM nonetheless gives a useful quantitative phenomenology for the dielectric properties of both pure \cite{hemberger95,hemberger96} and doped\cite{kleemann00,kleemann02,kleemann98_di,wu03,guo12} SrTiO$_3$.

The local nature of the Ising pseudospins makes the TIM valuable as a model for inhomogeneous systems, including doped quantum paraelectrics \cite{kleemann00,kleemann02,kleemann98_di,wu03,guo12}, ferroelectric thin films \cite{wangcl92,sun08,oubelkacem09,wangCD10,lu13,li16}, superlattices \cite{wangCL00,yao02}, and various low-dimensional structures \cite{xin99,lang07,lu14}. However, we show here that the TIM, as it is conventionally formulated, fails to correctly describe SrTiO$_3$ whenever nanoscale inhomogeneity is important. Most egregiously, the TIM fails to predict the formation of a quantized two-dimensional electron gas (2DEG) at LaAlO$_3$/SrTiO$_3$ interfaces, in contradiction with both theory and experiments \cite{gariglio15}. The goal of this paper is to propose a modification that we believe captures the essential physics of spatial inhomogeneity, and to compare it to the conventional TIM for model SrTiO$_3$ thin films and interfaces.
%The transverse Ising model (TIM) is a simple microscopic model that was developed to describe the ferroelectric transition in hydrogen-bonded materials like potassium dihydrogen phosphate (KDP) \cite{degennes63}. Because the TIM starts from a picture of fluctuating local electric dipole moments, it naturally describes materials, like KDP, with order-disorder transitions. Furthermore, mean-field formulations of the model have also been widely applied to displacive ferroelectrics, where they provide an especially useful phenomenology for inhomogeneous systems, including doped quantum paraelectrics \cite{kleemann00,kleemann02,kleemann98_di,wu03,guo12}, ferroelectric thin films \cite{wangcl92,sun08,oubelkacem09,wangCD10,lu13,li16}, superlattices \cite{wangCL00,yao02}, and various low-dimensional structures \cite{xin99,lang07,lu14}. However, we show here that the TIM, as it is conventionally formulated, fails to correctly describe SrTiO$_3$ whenever nanoscale inhomogeneity is important. Most egregiously, the TIM fails to predict the formation of a quantized two-dimensional electron gas (2DEG) at LaAlO$_3$/SrTiO$_3$ interfaces, in contradiction with both theory and experiments \cite{gariglio15}. The goal of this paper is to propose a modification that we believe captures the essential physics of spatial inhomogeneity, and to compare it to the conventional TIM for model SrTiO$_3$ thin films and interfaces.

In the TIM, the lattice polarization $P_{i}$ in unit cell $i$ is modelled by a pseudospin. This polarization is given by
\begin{equation} \label{P}
P_{i} = \mu \eta S^{(3)}_{i},
\end{equation}
where $\mu$ sets the scale of the electric dipole moment, $\eta = a^{-3}$ is the volume density of dipoles, and $a$ is the lattice constant. The pseudospin is usually taken to be $S=\frac 12$, and $S^{(3)}_{i}$ is the third component of the corresponding three-dimensional pseudospin vector $\bt{S}_{i}$. The other two components, $S^{(1)}_{i}$ and $S^{(2)}_{i}$, are fictitious degrees of freedom, with only the projection of ${\bf S}_i$ onto the $(3)$-axis corresponding to the physical polarization. (The unpolarized state is therefore described by the pseudospin lying entirely in the $(1)$-$(2)$ plane.) In a quantum model, $S^{(3)}_{i}$ is the expectation value of the operator $\hat{S}^{(3)}_{i}$, which is identical to the spin matrix $\hat{S}^{z}$ but which acts within pseudospin space.

The simplest version of the $S = \frac 12$ TIM is \cite{hemberger96}
\begin{equation} \label{TIM_orig}
\hat{H} = - \Omega \sum_{i} \hat{S}^{(1)}_{i} - J_{1} \sum_{\langle i, i' \rangle} \hat{S}^{(3)}_{i} \hat{S}^{(3)}_{i'} - \mu \sum_{i} E_{i} \hat{S}^{(3)}_{i},
\end{equation}
where $\Omega$ plays the role of a transverse magnetic field that flips the Ising spins, $J_1$ is a nearest-neighbour coupling constant with $\langle i,i' \rangle$ indicating nearest-neighbour sites, and $E_i$ is the electric field in unit cell $i$. For $J_1>0$, the model tends towards a ferroelectric state at low temperatures; however, this is limited by $\Omega$, which disorders the ferroelectric state. Under mean-field theory the model predicts a ferroelectric phase transition only if $\Omega < Z J_{1}$, where $Z$ is the coordination number of the lattice.

Although the TIM is only microscopically justified for order-disorder ferroelectrics, it is often used as a tool to characterize ferroelectrics of all types, and variations of this model have been applied to ferroelectricity in perovskites, including BaTiO$_{3}$ \cite{zhang00} and SrTiO$_3$ (STO) \cite{hemberger96}. As a phenomenological model, the TIM is more complex than simple Landau-Ginzburg-Devonshire theories; however, it is also more versatile.  The TIM, for example, is particularly well-suited to doped quantum paraelectrics, namely Sr$_{1-x}$M$_x$TiO$_3$ with M typically representing Ca or Ba \cite{kleemann00,kleemann02,kleemann98_di,wu03,tao04,guo12}. In these materials, small dopant concentrations are sufficient to induce a ferroelectric transition. Several groups have successfully modeled these materials as binary alloys of SrTiO$_3$ and MTiO$_3$ with doping-independent model parameters \cite{kleemann02,kleemann98_di,wu03,tao04,guo12}.

The current work is motivated by the application of the TIM to  metallic LaAlO$_{3}$/SrTiO$_{3}$ (LAO/STO) interfaces. These, and other related perovskite interfaces, have been widely studied since the discovery in 2004 that a 2DEG appears spontaneously at the interface when the LAO film is more than four unit cells thick \cite{ohtomo04}. This system is rich with interesting properties, including coexisting ferromagnetism and superconductivity \cite{Brinkman:2007fk,Reyren:2007gv,Dikin:2011gl}, nontrivial spin-orbit effects \cite{BenShalom:2010kv,Caviglia:2010jv}, a metal-insulator transition \cite{thiel06,Liao:2011bk}, gate-controlled superconductivity \cite{Caviglia:2008uh}, and a possible nematic transition at (111) interfaces \cite{Miao:2016hr,Davis:2017,Boudjada:2018,Boudjada:2019}. Furthermore, STO's proximity to the ferroelectric state has led to suggestions that quantum fluctuations shape its band structure \cite{atkinson17} and support  superconductivity \cite{Edge:2015fj,Dunnett:2018}. More generally, there has been a growing appreciation that lattice degrees of freedom play a key role in shaping the electronic structure near LAO/STO interfaces \cite{Behtash:2016dt,Lee:2016dj,Gazquez:2017bu,raslan18}.  With this in mind, the recent discovery that ferroelectric-like properties persist in some metallic perovskites \cite{Rischau:2017vj} naturally leads  one to explore the effects of Ca or Ba doping on LAO/STO interfaces and, as described above, the TIM provides a natural framework in which to do this.

We found, however, that the TIM as it is usually formulated in equation~\eref{TIM_orig} cannot reproduce the interfacial 2DEG and therefore fails to describe even the simple LAO/STO interface. In this work, we explain the reason for this failure and propose a modification to the TIM. In \sref{sec:FE}, we introduce the modified model and by comparison with the standard Landau-Ginzburg-Devonshire (LGD) expansion, illustrate why the failure arises and how we fix it. As a simple example, we apply the modified model to ferroelectric thin films.  In \sref{sec:interface}, we then apply the model to the LAO/STO interface, and show explicitly how the modification allows for the formation of the 2DEG.

\section{Inhomogeneous Ferroelectrics}
\label{sec:FE}

We begin by describing a modified TIM (\sref{sec:TIM}) that contains an additional anisotropic interaction; depending on its sign, this interaction generates either a pseudospin easy axis or easy plane. We obtain mean-field equations for the pseudospin and susceptibility, and by comparison to the LGD theory (\sref{sec:LGD}) we show that the Landau parameters are under-determined by the conventional TIM. Essentially, the problem is that equation~\eref{TIM_orig} contains three adjustable parameters ($\Omega$, $J_1$, and $\mu$), while the simplest LGD model requires four parameters to describe an inhomogeneous system. The additional interaction in the modified TIM fixes this discrepancy. In sections~\ref{sec:fit} and \ref{sec:J1} we obtain fits to the model parameters for the case of STO. As a simple application, in \sref{sec:FEfilm} we explore how the new term modifies the polarization distribution of a ferroelectric thin film.

\subsection{The Modified TIM}
\label{sec:TIM}

The modified Hamiltonian for general pseudospin $S$ is
\begin{eqnarray} \label{TIM_full}
\fl \hat{H} = - \Omega \sum_{i} \hat{S}^{(1)}_{i} - \frac{J_{1}}{2S} \sum_{\langle i,i' \rangle} \hat{S}^{(3)}_{i} \hat{S}^{(3)}_{i'} - \frac{J_\mathrm{an}}{2S} \sum_{i} \hat{S}^{(3)}_{i} \hat{S}^{(3)}_{i} - \mu \sum_{i} E_{i} \hat{S}^{(3)}_{i}.
\end{eqnarray}
This is equivalent to the Blume-Capel model in a transverse magnetic field \cite{Albayrak:2013}. The third term introduces an anisotropic pseudospin energy. If $J_\mathrm{an} > 0$, this term tends to align dipoles along the (3)-axis, making it an easy axis, which enhances the polarization; if $J_\mathrm{an} < 0$, the term tilts the dipole away from the (3)-axis, creating an easy plane and reducing the polarization.

The TIM is traditionally formulated with a spin-$\frac 12$ pseudospin. In that case, $\hat{S}^{(3)}_{i}$ is written in terms of a Pauli spin matrix, and $(\hat{S}^{(3)}_{i})^2$ is proportional to the identity operator. The new term therefore does not produce the desired anisotropy when $S = \frac{1}{2}$. This problem does not exist for higher spin models, and for this reason we formulate the TIM in terms of a general pseudospin $S$. However, we will show below that at the mean-field level, the model provides nearly the same results for any value of $S$, and for simplicity we revert to $S=1$ when we show results as a way of gaining insight into the general case.

Applying mean-field theory to equation~\eref{TIM_full} gives the following self-consistent expression for $S^{(3)}_{i}$:
\begin{equation} \label{S3}
S^{(3)}_i = \frac{S h^{(3)}_i}{h_i} f_{S}(h_i),
\end{equation}
where
\begin{equation} \label{f_S}
f_{S}(h_i) = \frac{1}{S} \frac{\sum\limits_{l=-S}^{S} l \rme^{\beta h_i l}}{\sum\limits_{n = -S}^{S} \rme^{\beta h_i n}} = B_S(\beta h_i S),
\end{equation}
$B_S(x)$ is the Brillouin function, $\beta = (k_\mathrm{B} T)^{-1}$, $T$ is temperature, $h_{i} = | \bt{h}_{i} |$, and $h_i^{(3)}$ is the $(3)$-component of the Weiss mean field for lattice site $i$,
\begin{equation} \label{h_i}
\textbf{h}_i = \left( \Omega, 0, \frac{J_{1}}{S} \sum_{i'} S^{(3)}_{i'} + \frac{J_\mathrm{an}}{S} S^{(3)}_i + \mu E_{i} \right).
\end{equation}
The summation $\sum_{i'}$ is a sum over the nearest neighbours of site $i$, and therefore depends on whether pseudospin $i$ is in a surface or bulk layer.

We linearize equation~\eref{S3} to obtain the condition that ensures ferroelectricity. In the uniform case,
\begin{equation} \label{h_uniform}
\bt{h} = \left( \Omega, 0, \frac{J_{0}}{S} S^{(3)} + \mu E \right),
\end{equation}
where
\begin{equation} \label{J0}
J_0 = ZJ_1 + J_\mathrm{an},
\end{equation}
for coordination number $Z$. At zero-temperature, $f_{S}(h_{i}) \rightarrow 1$, and from equation~\eref{S3} the model therefore predicts a paraelectric-ferroelectric phase transition when
\begin{equation}
S^{(3)} = \frac{J_0S^{(3)}}{\Omega}.
\end{equation}
From this one sees that, for any $S$,  a ferroelectric transition occurs at nonzero temperature only when $J_0 > \Omega$. In the case of a paraelectric like STO, $J_0 < \Omega$.

\begin{figure}[]
	\centering
	\includegraphics[width=0.7\linewidth]{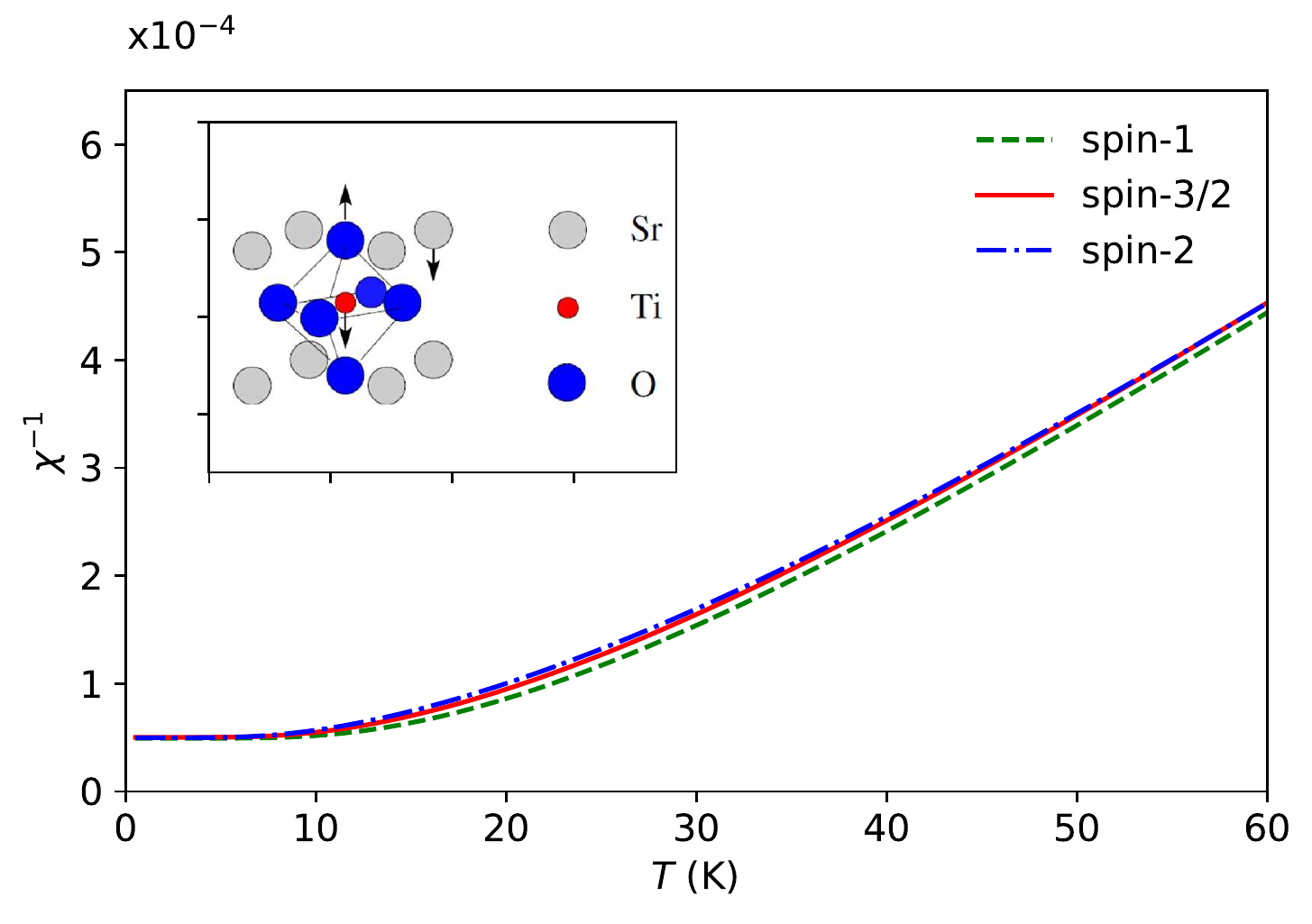}
	\caption{Inverse dielectric susceptibility versus temperature for SrTiO$_{3}$, modelled using three-, four- and five-component pseudospins. The fitting parameters were found separately for each pseudospin (\tref{tab:pars}). {\it{Inset}}: The SrTiO$_3$ unit cell is illustrated, showing that the polarization is primarily due to the soft phonon mode (black arrows), in which the oxygen cage moves opposite to the titanium ions. The inset is re-published from \cite{atkinson17}.}
	\label{fig:STOX_compspin}
\end{figure}

To show that the choice of $S$ has a small effect at the mean-field level, the uniform inverse dielectric susceptibility of STO is plotted for different values of $S$ in \fref{fig:STOX_compspin}. From equation~\eref{P}, the susceptibility for a weak uniform electric field $E$ is 
\begin{equation} \label{X_gen}
\chi (T) = \left . \frac{1}{\epsilon_0} \frac{dP}{dE} \right|_{E=0} =\left . \frac{\mu \eta}{\epsilon_{0}} \frac{d S^{(3)}}{dE} \right|_{E=0},
\end{equation}
where $dS^{(3)}/dE$ is obtained from equation~\eref{S3} with $\bt{h}$ given by equation~\eref{h_uniform}. \Fref{fig:STOX_compspin} shows results for $S=1$, $S=\frac 32$ and $S=2$. The fitting parameters $J_0$, $\mu$, and $\Omega$ depend on the value of $S$ and were determined by fitting to the experimental susceptibility, as described in \sref{sec:fit} below. (Note that $J_1$ is not explicitly used here because the calculations are for bulk STO.) The values of all these parameters are listed in \tref{tab:pars}.

Because the model was fitted to low- and high-temperature susceptibilities, the curves in \fref{fig:STOX_compspin} are expected to be close in value at these limits. However, they also differ only slightly in between, indicating that STO is well-described by the simplest case shown, $S=1$, when using mean-field theory. In particular, the model accurately captures both Curie-Weiss behaviour at high temperature, and the saturation of the susceptibility at low temperature (where the ferroelectric transition is suppressed by quantum fluctuations).

\begin{table}
	\caption{Model parameters for SrTiO$_3$. Parameters are obtained from fits to the experimental susceptibility~\cite{sakudo71} and phonon dispersion \cite{cowley64}.}
	\centering
	\begin{indented}
		\item[]\begin{tabular}{c c c c}
			\br
			&  Spin-1 & Spin-3/2 & Spin-2 \\
			\mr
			$\Omega$ (meV) & 4.41 & 3.53 & 2.94 \\
			$J_{0}$ (meV) & 3.88 & 3.10 & 2.58 \\
			$J_{1}$ (meV) & 30-130 & 40-160 & 50-200 \\
			$\mu$ ($e$\AA) & 1.88 & 1.37 & 1.09 \\
			\br
		\end{tabular}
	\end{indented}
	\label{tab:pars}
\end{table}

\subsection{Comparison to the Landau-Ginzburg-Devonshire Expansion}
\label{sec:LGD}

While equation~\eref{S3} is the fundamental self-consistent equation for $S^{(3)}_{i}$, the role each parameter plays in determining the pseudospin is not transparent. For example, it is not immediately evident from this expression why the conventional TIM  (with $J_\mathrm{an}=0$) is unable to describe inhomogeneous systems. To explore this point, we expand equation~\eref{S3} in powers of $h_{i}^{(3)}$ and compare the coefficients to those in a typical LGD expansion. We show that the transition temperature and correlation length cannot be set independently unless $J_\mathrm{an}$ is nonzero.

The typical LGD free energy with order parameter $S^{(3)}(\bt{r})$ has the form
\begin{eqnarray} \label{F}
\mathcal{F} &=& \eta \int d^{3} r\,  \Bigg[ \frac{A}{2} \left( S^{(3)}(\bt{r}) \right)^{2} + \frac{B}{4}  \left( S^{(3)}(\bt{r}) \right)^{4} \nonumber \\ && + \frac{C}{2} \left( \nabla S^{(3)}(\bt{r}) \right)^{2} - D E(\bt{r}) S^{(3)}(\bt{r}) \Bigg].
\end{eqnarray}
$E(\bt{r})$ is the electric field, $A$, $B$, $C$ and $D$ are the LGD coefficients that describe the material, and $\eta$ is the inverse volume of a unit cell. Minimizing equation~\eref{F} with respect to $S^{(3)}(\bt{r})$ gives the familiar equation
\begin{equation} \label{Fmin}
0 = A S^{(3)}(\bt{r}) + B \left( S^{(3)}(\bt{r})\right)^{3} -C \nabla^{2} S^{(3)}(\bt{r}) - D E(\bt{r}),
\end{equation}
which can be solved for the pseudospin. The critical temperature is set by $A$, which changes sign at the ferroelectric transition, while $B$ determines the zero-temperature polarization. In the paraelectric phase, $D$ is determined by the dielectric susceptibility and $C$ and $A$ set the correlation length $\xi=\sqrt{C/A}$.

We expand equation~\eref{S3} in powers of $h_{i}^{(3)}$ to obtain
\begin{equation} \label{expand_2}
S^{(3)}_i = \frac{S f_S (\Omega)}{\Omega} h^{(3)}_i + \frac{1}{2 \Omega} \left( \frac{d}{d\Omega} \frac{Sf_{S}(\Omega)}{\Omega} \right) \left( h^{(3)}_{i} \right)^{3},
\end{equation}
where ${h}_i$ and $h_i^{(3)}$ are defined by equation~\eref{h_i}. To proceed further, we note that the discretized second derivative of a  function $f_j=f(x_j)$ is
\begin{equation}
\left . \frac{d^{2}f(x) }{dx^{2}} \right |_{x=x_j} \approx \frac{f_{j-1} - 2f_j + f_{j+1}}{a^{2}}.
\end{equation}
Then, equation~\eref{h_i} can be re-written as
\begin{equation}
h^{(3)}_i = \frac{J_0}{S} S^{(3)}_i + \frac{J_1}{S} a^{2} \nabla^{2} S^{(3)}_i + \mu E_i,
\end{equation}
with $J_0$ defined by equation~\eref{J0}. This can now be substituted into equation~\eref{expand_2}.

Keeping only terms that are directly comparable to those in equation~\eref{Fmin}, we obtain
\numparts
\begin{eqnarray}
A & = \frac{\Omega}{Sf_{S}(\Omega)} - \frac{J_{0}}{S}, \label{coeff_A} \\
B & = -\frac{1}{2 S f_{S}(\Omega)} \left( \frac{J_{0}}{S} \right)^{3} \frac{d}{d\Omega} \left( \frac{S f_S (\Omega)}{\Omega} \right), \label{coeff_B} \\
C & = \frac{J_{1} a^{2}}{S}, \label{coeff_C} \\
D & = \mu. \label{coeff_D}
\end{eqnarray}
\endnumparts
These equations show that $A$ and $B$ are determined by combinations of $J_0$ and $\Omega$, while $C$ and $D$ are determined by $J_1$ and $\mu$, respectively. The key point is that $J_0$ reduces to $ZJ_1$ for the conventional TIM, in which case $A$ and $C$ are not independent. Physically, this means that the correlation length, which sets the length scale over which the material responds to inhomogeneities, cannot be determined independently of the transition temperature and low-$T$ polarization. In other words, the four coefficients $A$, $B$, $C$ and $D$ are only described by three parameters, $\Omega$, $J_{1}$ and $\mu$.

In this case, the model predicts a significantly smaller correlation length at low temperatures than does the modified TIM. From equations~(\ref{coeff_A}) and (\ref{coeff_C}),
\begin{equation}
\xi = \sqrt{\frac{C}{A}} = \sqrt{\frac{J_1 a^2}{\frac{\Omega}{f_S(h)} - J_0}}.
\end{equation}
At low temperatures, $f_S(\Omega) \rightarrow 1$. In this case, the conventional TIM ($J_1 = J_0/Z$) gives $\xi\approx4.3$~\AA, independent of $S$. For $S=1$, the range of correlation lengths from the modified TIM, where the $J_1$ values are taken from \tref{tab:pars}, is 2.9-6.1 nm, which is an order of magnitude larger. The pseudospin anisotropy $J_\mathrm{an}$ is therefore an essential part of the TIM.

\subsection{Fitting $\Omega$, $J_0$, and $\mu$ for SrTiO$_3$}
\label{sec:fit}

Most of the TIM parameters can be fit to existing susceptibility data. We do this for STO, as it will form the basis of our discussion in \sref{sec:interface}. 

Inserting equation~\eref{S3} into equation~\eref{X_gen}, we obtain the susceptibility
\begin{equation} \label{X}
\chi(T,0) = \frac{\mu^{2} \eta}{\epsilon_{0}} \frac{1}{L(h,T) - J_{0}/S} \Bigg\vert_{E=0},
\end{equation}
where $h = |\bt{h}|$, $\bt{h}$ is given by equation~\eref{h_uniform}, and
\begin{eqnarray}
\fl L(h,T) = \Bigg[ S \left( \frac{1}{h} - \frac{\left( h^{(3)} \right)^{2}}{h^{3}} \right) f_S(h) + S \frac{\left( h^{(3)} \right)^{2}}{h^{2}} \frac{\partial f_S(h)}{\partial h} \Bigg]^{-1}.
\end{eqnarray}

At high temperatures, this expression simplifies. Taking $L(h,T)|_{T \rightarrow \infty}=[\beta S(S+1)/3]^{-1}$, equation~\eref{X} obtains a Curie-Weiss form, 
\begin{equation} \label{X_0}
\chi(T,0) = \frac{\mu^{2} \eta S(S+1) }{3 \epsilon_{0} k_B} \frac{1}{T - T_\mathrm{CW}},
\end{equation}
where $T_\mathrm{CW} = (S+1) J_{0}/3k_\mathrm{B} \approx 30$~K \cite{sakudo71} is the transition temperature implied by the high-temperature susceptibility. (In STO, this transition is suppressed by quantum fluctuations.) $J_0$ and $\mu$ are thus obtained by matching equation~\eref{X_0} to high-$T$ experiments.

At low $T$, equation~\eref{X} takes one of two forms depending on whether the system is ferroelectric or not. For a ferroelectric, $\Omega$ can be found from the behaviour of the susceptibility at $T \rightarrow T_\mathrm{c}^{+}$ for critical temperature $T_\mathrm{c}$. In this case, $h^{(3)}=0$, $h=\Omega$, and setting the denominator of equation~\eref{X} to zero gives a self-consistent equation for $\Omega$,
\begin{equation}
\Omega = J_0 f_S(\Omega).
\end{equation}

For a paraelectric like STO, on the other hand, we obtain $\Omega$ from the zero-temperature susceptibility. In this limit $L(h,T)|_{T \rightarrow 0} = \Omega/S$ and equation~\eref{X} may easily be inverted for $\Omega$. The values of $J_{0}$, $\mu$ and $\Omega$ for STO determined from equation~\eref{X} are listed in \tref{tab:pars}.

The closeness in value between $J_0$ and $\Omega$ for STO can be understood from their physical meanings. $J_0$ sets the temperature at which a transition would occur in the absence of quantum fluctuations, while $\Omega$ sets the scale of the quantum fluctuations; that these two are close in value is because STO is close to a ferroelectric transition. Further, since the Curie-Weiss temperature is small, both of these parameters are small.

\subsection{Estimating $J_{1}$ for SrTiO$_{3}$}
\label{sec:J1}

As was shown in \sref{sec:LGD}, $J_{1}$ sets the scale of the gradient term $C$ in the LGD expansion, and it can therefore  be obtained from quantities related to spatial gradients of the polarization. In perovskites, the polarization is closely connected to an optical phonon mode \cite{cowley64,atkinson17}, pictured in \fref{fig:STOX_compspin}. One can therefore obtain $J_1$ from the phonon dispersion.

Key to this analysis is that the optical phonon has a large dipole moment that is represented by the TIM pseudospins. The phonon spectrum can therefore be obtained from the dynamical pseudospin correlation function. In the paraelectric phase, the term proportional to $\Omega$ in equation~\eref{TIM_full} ensures that the pseudospins lie primarily along the (1)-axis. Perturbations of this state can be viewed as the magnons of a fictitious ferromagnetic material in which the magnetic moments align along the (1)-axis. The phonons can then be described as spin-wave excitations.

The spin operators are difficult to work with, however, and it is useful to bosonize them. This is achieved with the Holstein-Primakoff transformation \cite{holstein,holstein40}. This transformation maps the pseudospin operators on to the boson creation and annihilation operators, $\hat{a}^{\dag}_{i}$ and $\hat{a}_{i}$. Pseudospin projections on the (3)-axis are then modelled as boson excitations, with a pseudospin that is entirely polarized along the (1)-axis represented by the vacuum state.

In this representation, the raising and lowering operators for site $i$ differ from the typical set by a cyclic permutation of the pseudospin axes. We then define \cite{holstein}
\numparts
\begin{eqnarray} \label{S+}
\hat{S}^{+}_{i} & = \hat{S}^{(2)}_{i} + i \hat{S}^{(3)}_{i} \\
& = \sqrt{2S} \left( 1 - \frac{1}{2S} \hat{a}^{\dag}_{i} \hat{a}_{i} \right)^{1/2} \ \hat{a}_{i},
\end{eqnarray}
\endnumparts
\numparts
\begin{eqnarray} \label{S-}
\hat{S}^{-}_{i} & = \hat{S}^{(2)}_{i} - i \hat{S}^{(3)}_{i} \\ 
& = \sqrt{2S} \hat{a}^{\dag}_{i} \left( 1 - \frac{1}{2S} \hat{a}^{\dag}_{i} \hat{a}_{i} \right)^{1/2}.
\end{eqnarray}
\endnumparts
Since the polarization lies close to the (1)-axis in the paraelectric state, only low bosonic excitation states are relevant. In this case, $\hat{S}^{+}_{i} \approx \sqrt{2S} \hat{a}_{i}$ and $\hat{S}^{-}_{i} \approx \sqrt{2S} \hat{a}^{\dag}_{i}$. Additionally, the (1)-component of the pseudospin is defined as \cite{holstein}
\begin{equation} \label{S1}
\hat{S}^{(1)}_{i} = S - \hat{a}^{\dag}_{i} \hat{a}_{i},
\end{equation}
and the (3)-component is
\begin{eqnarray} \label{S3_a}
\hat{S}^{(3)}_{i} & = \frac{1}{2i} \left( \hat{S}^{+}_{i} - \hat{S}^{-}_{i} \right) \\
& = \frac{\sqrt{2S}}{2i} \left( \hat{a}_{i} - \hat{a}^{\dag}_{i} \right).
\end{eqnarray}
Because $\hat{S}^{(3)}_{i}$ represents atomic displacements, $\hat{a}_i$ and $\hat{a}^{\dag}_i$ are therefore phonon operators.

Equations~\eref{S1} and \eref{S3_a} can now be substituted into equation~\eref{TIM_full}. We transform to reciprocal space using $\hat{a}_{i}$~=~$\frac{1}{\sqrt{N}}\sum_{\bt{k}}  e^{i\bt{k} \cdot \bt{r}_{i}} \hat{b}_{\bt{k}}$:
\begin{eqnarray}
\fl \hat{H} = - N \left( \Omega S + \frac{J_\mathrm{an}}{4} \right) + \sum_{\bt{k}} \Gamma_{\bt{k}} \hat{b}^{\dag}_{\bt{k}} \hat{b}_{\bt{k}} + \sum_{\bt{k}} \frac{\Delta_{\bt{k}}}{2} \left( \hat{b}_{\bt{k}} \hat{b}_{-\bt{k}} + \hat{b}^{\dag}_{\bt{k}} \hat{b}^{\dag}_{-\bt{k}} \right),
\end{eqnarray}
where $\gamma_{\bt{k}} = 2 \cos (k_{x}a) + 2 \cos (k_{y}a) + 2 \cos (k_{z} a)$, $N$ is the total number of lattice sites, and
\numparts
\begin{eqnarray}
\Delta_{\bt{k}} & = \frac{J_{1}}{2} \gamma_{\bt{k}} + \frac{J_\mathrm{an}}{2}, \\
\Gamma_{\bt{k}} & = \Omega - \Delta_{\bt{k}}.
\end{eqnarray}
\endnumparts
Note that we have set $E = 0$ here, since the phonon spectrum is measured at zero field.

It is convenient to formulate the dynamics of the pseudomagnons using Green's functions. The Green's functions are correlation functions between the pseudomagnon creation and annihilation operators, and the equations of motion of the Green's functions therefore include the equations of motion of $\hat{b}_{\bt{k}}$ and $\hat{b}^{\dag}_{\bt{k}}$. The spin-wave excitation spectrum can then be obtained from the poles of the Green's function.

The Green's function and its equation of motion are, respectively,
\begin{equation} \label{G1}
D_{1}(\bt{k}, t) = -i \left\langle \left[ \hat{b}_{\bt{k}}(t), \hat{b}^{\dag}_{\bt{k}}(0) \right] \right\rangle \theta (t),
\end{equation}
\begin{equation} \label{G1_eom}
\frac{dD_{1}(\bt{k}, t)}{dt} = -i \delta (t) - i \Gamma_{\bt{k}} D_{1}(\bt{k}, t) - i \Delta_{\bt{k}} D_{2}(\bt{k}, t),
\end{equation}
where $\theta(t)$ is the step function. The second Green's function that appears in equation~\eref{G1_eom} and its equation of motion are
\begin{equation} \label{G2}
D_{2}(\bt{k}, t) = -i \left\langle \left[ \hat{b}^{\dag}_{-\bt{k}} (t), \hat{b}^{\dag}_{\bt{k}} (0) \right] \right\rangle \theta (t),
\end{equation}
\begin{equation} \label{G2_eom}
\frac{dD_{2}(\bt{k}, t)}{dt} = i \Gamma_{-\bt{k}} D_{2}(\bt{k}, t) + i \Delta_{\bt{k}} D_{1}(\bt{k}, t).
\end{equation}
Fourier transforming equations~\eref{G1_eom} and \eref{G2_eom} in time and solving for $D_{1}(\bt{k}, \omega_{\bt{k}})$ gives the following expression for the Green's function:
\begin{equation} \label{G(om)}
D_{1}(\bt{k}, \omega_{\bt{k}}) = \frac{\omega_{\bt{k}} + \Gamma_{\bt{k}}}{\omega^{2}_{\bt{k}} - \Gamma_{\bt{k}}^{2} + \Delta_{\bt{k}}^{2}}.
\end{equation}
The phonon dispersion is therefore given by
\numparts
\begin{eqnarray}
\omega_{\bt{k}} & = \sqrt{\Gamma_{\bt{k}}^{2} - \Delta_{\bt{k}}^{2}} \\
& = \sqrt{\Omega \left( \Omega - 2\Delta_{\bt{k}} \right)}.
\end{eqnarray}
\endnumparts

We obtain an expression for $J_{1}$ by comparing the frequency at $k_{x} = \pi/2$ and the zone centre:
\begin{equation} \label{J1}
J_{1} = \frac{\hbar^{2} \left( \omega^{2}_{\pi/2} - \omega^{2}_{0} \right)}{\Omega \left( \gamma_{0} - \gamma_{\pi/2} \right)},
\end{equation}
where the subscripts $\pi/2$ and 0 indicate $\bt{k}=(\pi/2,0,0)$ and $\bt{k}=(0,0,0)$, respectively. Since $\Omega$ is already known from bulk susceptibility data, $J_{1}$ can be estimated solely using the material's phonon dispersion. Using neutron scattering data from \cite{cowley64}, we obtained a range of $J_{1}$ values between 30 and 200~meV depending on $S$ and on how the fit was made. As will be shown in \sref{sec:interface}, these estimates are somewhat lower than the values required to produce a 2DEG at the LAO/STO interface, which is likely a limitation of the TIM. Nonetheless, this calculation shows that $J_{1}$ is orders of magnitude larger than the value $J_1 = J_0/Z$ that is implicit in the conventional TIM.

This large discrepancy between $J_1$ and $J_0$ is a key feature of STO, and that there is more than an order of magnitude difference between their values can be related to their different physical origins. Further, from equation~\eref{J0} it follows that $J_\mathrm{an}$ is not small; rather, it is negative and nearly cancels $ZJ_1$. $J_\mathrm{an}$ would however play less of a role in a material with a high transition temperature, where $J_1$ and $J_0$ would be closer in value.

\subsection{Ferroelectric Thin Films}
\label{sec:FEfilm}

We first model the polarization in ferroelectric thin films as a simple application of the modified TIM. A ferroelectric's properties can vary drastically between the bulk and thin-film forms, and the origins and applications of these differences have been increasingly studied in recent years \cite{setter06}. Ferroelectric thin films provide significant advantages in electronic devices such as increased efficiency in photovoltaic cells \cite{liu16,zenkevich14,kutes14} and decreased power usage in non-volatile memory storage \cite{muller11}.

We focus on weakly ferroelectric materials, like those obtained by doping STO with $^{18}$O, Ca, or Ba. We take $S=1$, and we thus fix the parameters $J_{0}=3.88$~meV and $\mu=1.88$~$e$\AA, which were determined in section~\ref{sec:fit} for STO. To obtain a ferroelectric transition, we take $\Omega=3.2$~meV, which yields a bulk transition temperature $T_\mathrm{c}\approx20$~K,
similar to what is observed in Sr$_{1-x}$Ca$_x$TiO$_3$. We treat $J_1$ as an adjustable parameter.

Thin films have a layered geometry that simplifies calculations. Taking each layer to be one unit cell thick, and assuming translational invariance within the $xy$-plane, the pseudospin, electric field, and polarization depend only on the layer index $i_{z}$ (instead of site $i$). Equation~\eref{S3} becomes
\begin{equation} \label{S3_i}
S^{(3)}_{i_z} = \frac{S h^{(3)}_{i_z}}{h_{i_z}} f_{S}(h_{i_z}),
\end{equation}
where $h_{i_z} = |\bt{h}_{i_z}|$ and the Weiss mean field is 
\begin{equation} \label{eq:hz}
\textbf{h}_{i_z} = \left( \Omega, 0, \frac{J_{1}}{S} \sum_{i'} S^{(3)}_{i'} + \frac{J_\mathrm{an}}{S} S^{(3)}_{i_z} + \mu E_{i_z} \right),
\end{equation}
where, for the cubic STO crystal structure, the sum over nearest neighbours of a pseudospin in layer $i_z$ is $\sum_{i'}S_{i'} = 4S^{(3)}_{i_z} + S^{(3)}_{i_z-1} + S^{(3)}_{i_z+1}$. The lattice polarization in layer $i_{z}$ is then
\begin{equation} \label{P_i}
P_{i_{z}} = \mu \eta S^{(3)}_{i_{z}}.
\end{equation}
(Recall that $\mu S$ is the maximum dipole moment per unit cell and $\eta$ is the dipole moment density.)

We assume a short-circuit geometry, in which the top and bottom surfaces of the film are connected by a wire that maintains a zero voltage difference between them. This geometry is commonly adopted to minimize the effects of depolarizing electric fields. We thus have two kinds of charge: a bound charge $\rho^\mathrm{b}(z) = - \partial_z P_\mathrm{tot}(z)$ coming from a sum of atomic and lattice polarizations, and the external charges $\rho^\mathrm{ext}(z)$ on the top and bottom electrodes.

The electric field in equation~\eref{eq:hz}  is obtained from these charges via Gauss' law, 
\begin{equation}
\epsilon_{0} \frac{d}{dz} E(z) = \rho^\mathrm{b}(z) + \rho^\mathrm{ext}(z).
\end{equation}
We break the polarization into lattice and atomic pieces, $P(z)$ and $\epsilon_0 \alpha E(z)$ respectively, with $\alpha$ the atomic polarizability, and defining the optical dielectric constant $\epsilon_\infty=\epsilon_0(1+\alpha)\approx 5.5 \epsilon_0$ \cite{raslan17,zollner00}, we obtain the usual expression 
\begin{equation} \label{Gauss}
\frac{d}{dz} \left[ \epsilon_{\infty} E(z) + P(z) \right] = \rho^\mathrm{ext}(z),
\end{equation}
which can be integrated to find $E(z)$.

The charge density in the top and bottom electrodes is written as 
\begin{equation}
\rho^\mathrm{ext}(z) = \frac{en}{a^{2}} [ \delta(z) - \delta(z-L) ],
\end{equation}
where $L$ is the film thickness, and $n$ is the positive charge per 2D unit cell on the top electrode. Integrating equation~\eref{Gauss} gives 
\begin{equation} \label{E_Gauss}
\epsilon_{\infty} E(z) = - P(z) + \frac{en}{a^{2}}.
\end{equation}
A second integration, of equation~\eref{E_Gauss} across the thickness of the film, gives
\begin{equation} \label{eq:ena2}
\frac{en}{a^{2}} = \frac{\int_{0}^{L}dz P(z) - \epsilon_{\infty} V}{L},
\end{equation}
with $V$ the potential difference across the film. Using this to eliminate $en/a^2$ in equation~\eref{E_Gauss}, and setting $V=0$ for the short-circuit geometry, we obtain
\begin{equation} \label{E}
E(z) = \frac{P_\mathrm{ave} - P(z)}{\epsilon_{\infty}},
\end{equation}
with $P_\mathrm{ave}$ the average polarization of the film. Equations~\eref{eq:hz} and \eref{E} are evaluated at discrete positions $z = i_z a$, and together with equation~\eref{S3_i} form a closed set that can be solved self-consistently.

\begin{figure}[tb]
	\includegraphics[width = 0.5\linewidth]{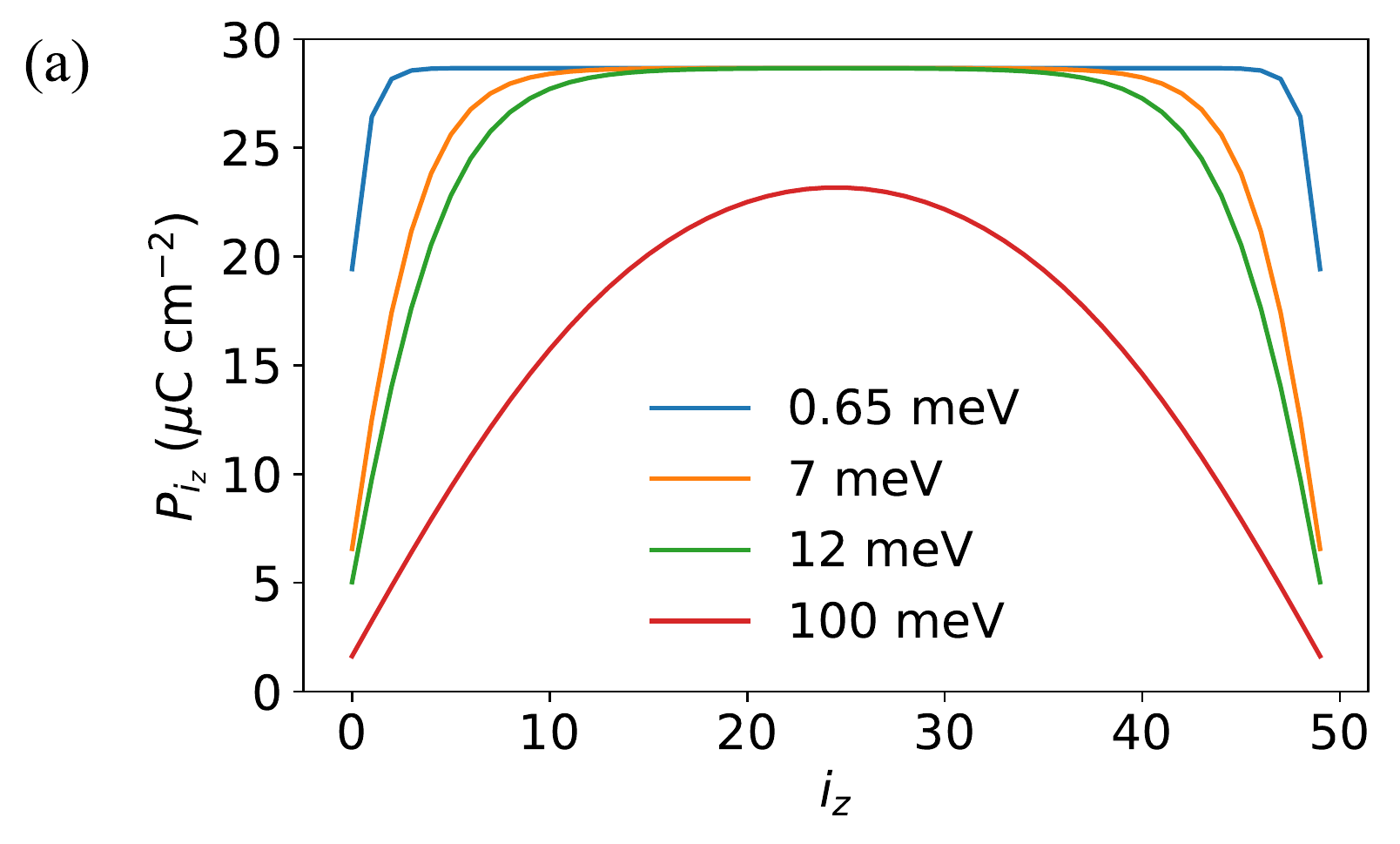}
	\includegraphics[width = 0.5\linewidth]{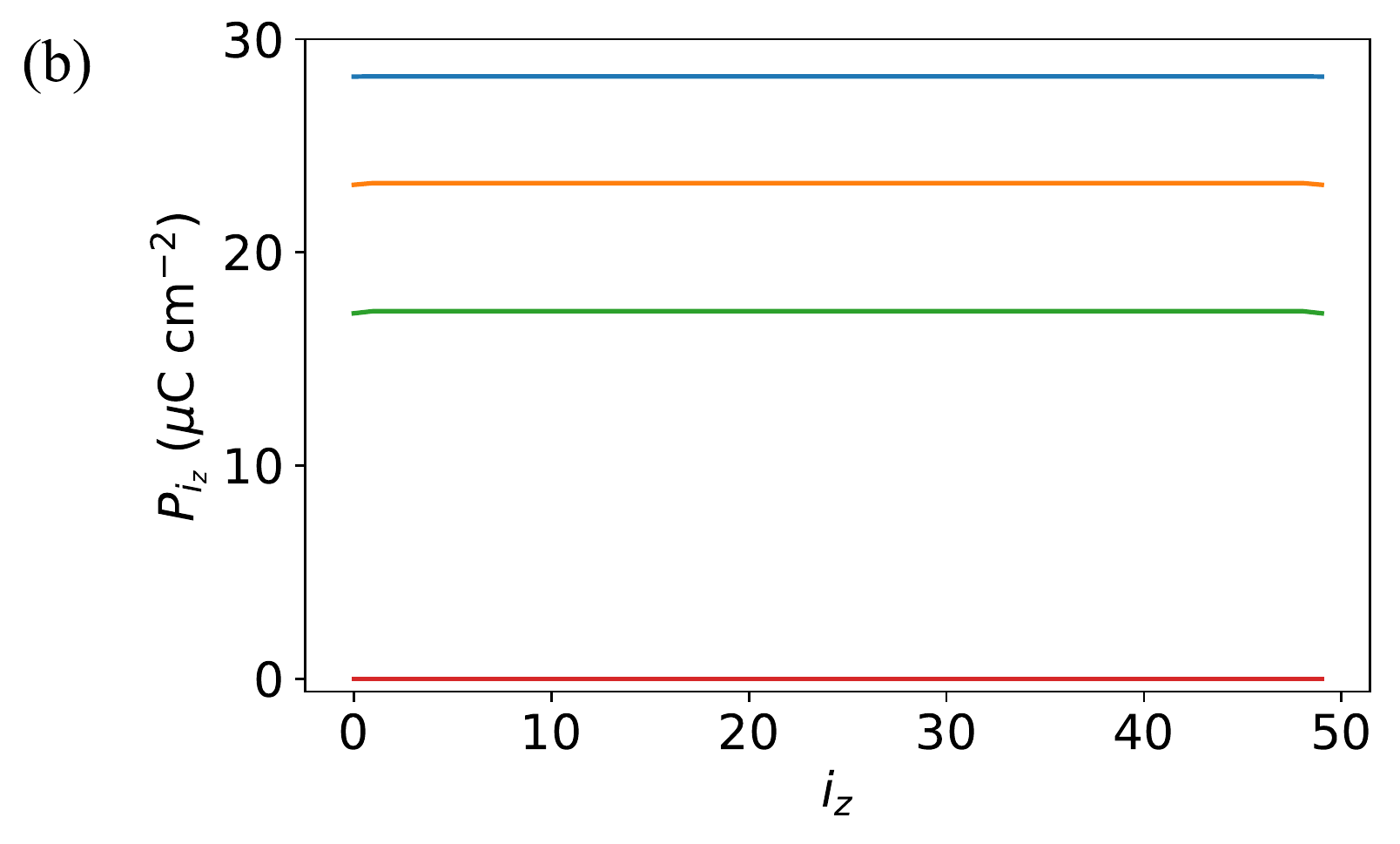}
	\caption{Polarization versus layer for 50-layer films (a) without and (b) with  the depolarizing field included. Results are shown for $J_1$ values between 0.65~meV and 100~meV, with $T$~=~1~K, $J_{0}=3.88$~meV, $\Omega =3.2$~meV and $\mu=1.88$~$e$\AA. $J_1=0.65$~meV corresponds to the conventional transverse Ising model ($J_\mathrm{an} = 0$). For reference, the bulk polarization is $P_\mathrm{bulk} = 29 \ \mu$C cm$^{-2}$. Here and throughout, results are shown for $S=1$.}
	\label{fig:Ecomp}
\end{figure}

\Fref{fig:Ecomp} shows the results of simulations for a film that is $N_L = 50$ layers thick. The figure illustrates two main points: First, the results depend qualitatively on whether or not electric fields are included in the simulation, even in the short-circuit geometry (for which naive considerations suggest the field vanishes); second, for fixed $J_0$, the value of $J_1$ has a large impact on the polarization.

The effects of electric fields in thin films were discussed at length by Kretschmer and Binder \cite{kretschmer79}, and the results in \fref{fig:Ecomp} serve as a reminder of their importance. In \fref{fig:Ecomp}(a), where electric fields are not included, the polarization is reduced at the surfaces and increases to its bulk value over a length scale set by the correlation length. In the ferroelectric phase, the correlation length is $\xi=\sqrt{-C/2A}$ (in terms of LGD parameters), which is proportional to $\sqrt{J_1}$. The conventional TIM with $J_\mathrm{an}=0$ has $J_1=0.65$~meV, which corresponds to a correlation length of $\xi=2.7$ \AA. Consistent with this, \fref{fig:Ecomp}(a) shows that for the conventional TIM, surface effects are confined to narrow regions near the edges of the film. Conversely, the modified TIM with a more realistic value of $J_1=100$~meV gives the correlation length $\xi=3.3$~nm, which is comparable to the film thickness. In this case, the polarization is inhomogeneous throughout the film. In contrast to both of these cases, the polarization is nearly constant across the film when electric fields are included [\fref{fig:Ecomp}(b)]; the polarization decreases with increasing $J_1$, and is suppressed completely for $J_1=100$~meV.

The apparent uniformity of the polarization across the film in \fref{fig:Ecomp}(b) is because the correlation length is replaced by a shorter length scale $\kappa^{-1}$ when electric fields are included, with \cite{kretschmer79}
\begin{equation} \label{kappa}
\kappa = \sqrt{\xi^{-2} + \frac{\mu^{2} \eta}{\epsilon_{0} C}}.
\end{equation}
In STO, this length scale is less than a unit cell, and the polarization is therefore nearly constant, with only a small reduction in the surface layer. This slight reduction is, nonetheless, enough that the depolarizing fields are incompletely screened by the electrodes. There is thus a residual depolarizing field in the STO film that reduces the overall polarization of the film.

\begin{figure*}[tb]
	\centering
	\includegraphics[width = 0.3\linewidth]{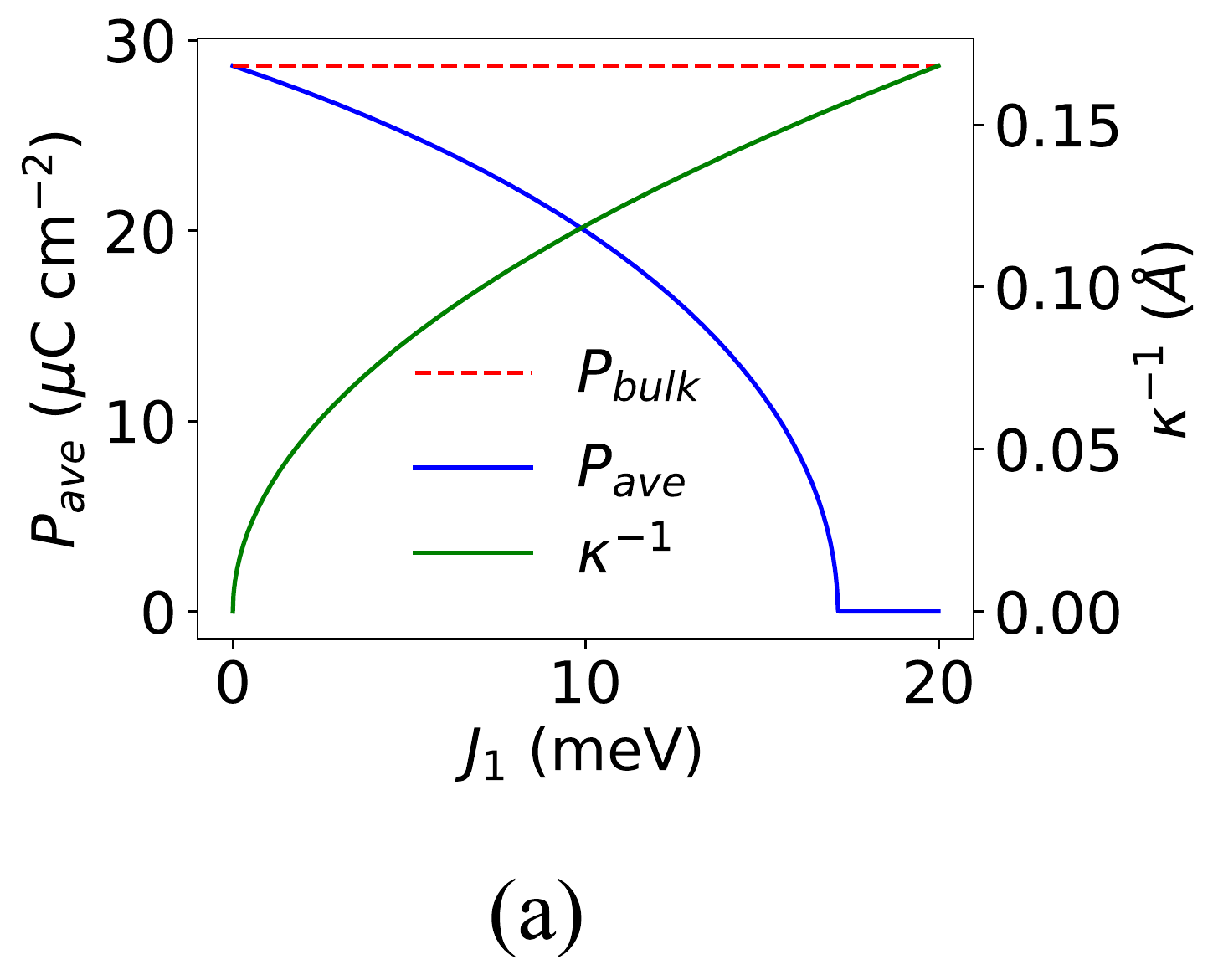}
	\includegraphics[width = 0.3\linewidth]{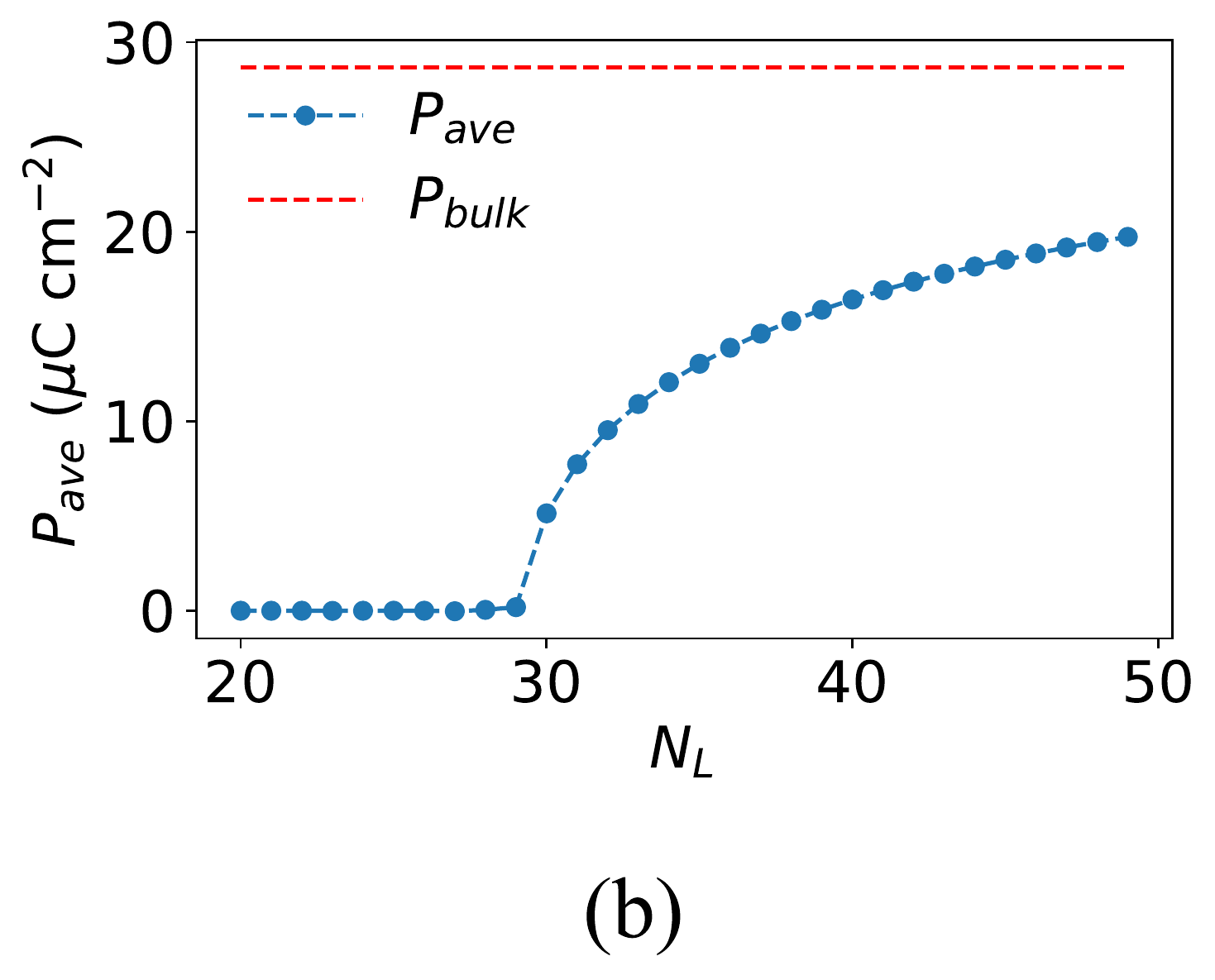}
	\includegraphics[width = 0.3\linewidth]{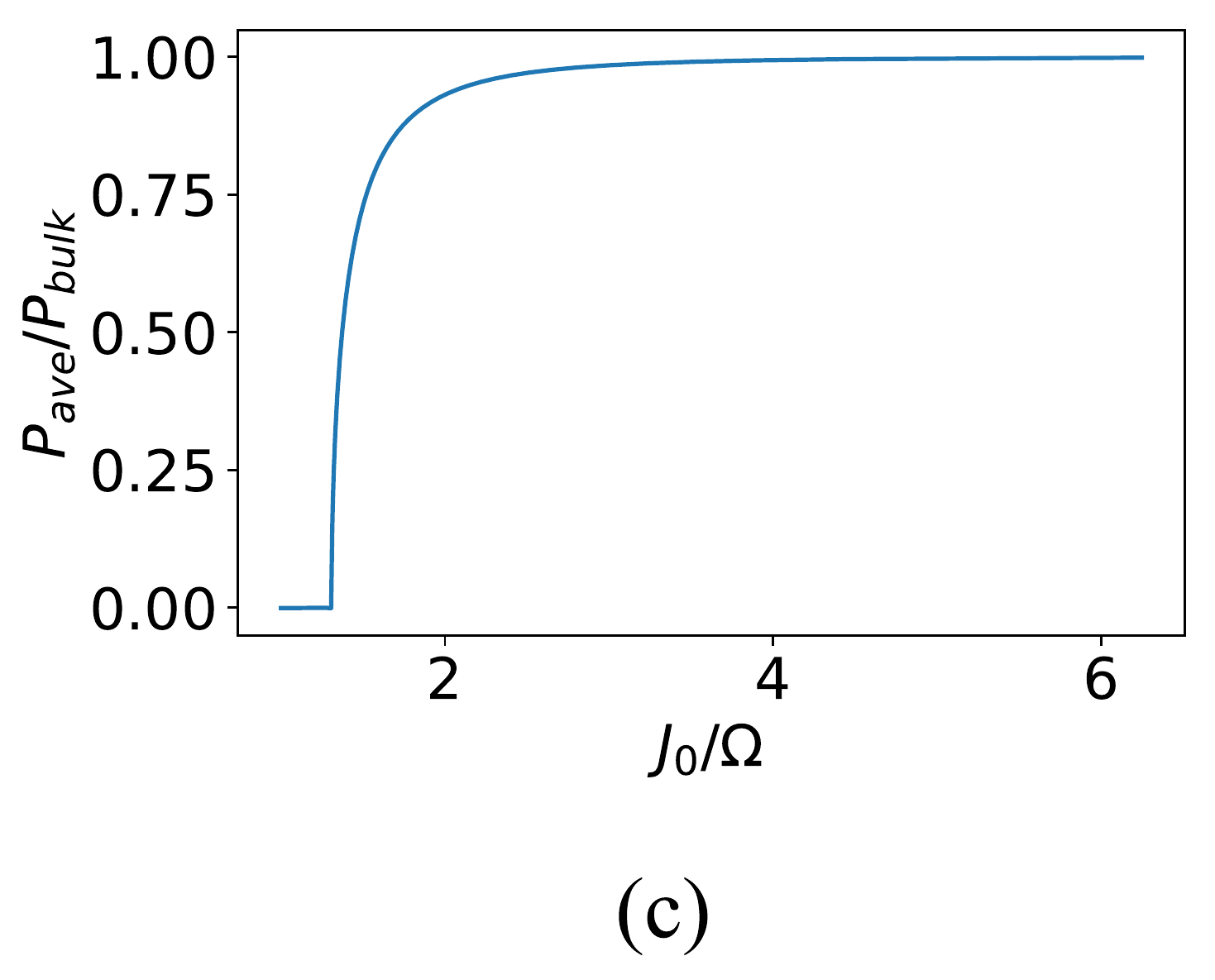}
	\caption{Average polarization $P_\mathrm{ave}$ of a ferroelectric thin film in the short-circuit geometry, including the depolarizing field. Because the polarization is nearly uniform, $P_\mathrm{ave}$ is almost the same as the polarization in each layer.	(a) Dependence of $P_\mathrm{ave}$ (blue) and $\kappa^{-1}$ (green) on $J_{1}$; also shown for comparison is the bulk value $P_\mathrm{bulk}$ of the polarization; (b) dependence of $P_\mathrm{ave}$ on film thickness $N_{L}$ (where $N_L=L/a$ is the total number of layers); (c) dependence of $P_\mathrm{ave}$ on $J_{0}$. Except where otherwise indicated, parameters are the same as in \fref{fig:Ecomp} and $N_L=50$. In (b), $J_1 = 10$~meV; in (c), $J_1 = 25$~meV.}
	\label{fig:Pdep}
\end{figure*}

To make the dependence of $\kappa$ on the TIM parameters explicit, we substitute values for the LGD parameters from equations~\eref{coeff_A}-\eref{coeff_D} into equation~\eref{kappa} in the limit $T\rightarrow 0$. For spin-1 we find
\begin{equation}
\kappa = \sqrt{-\frac{2(\Omega - J_{0})}{J_{1}a^{2}} + \frac{\mu^{2} \eta}{\epsilon_{0} J_{1} a^{2}}}.
\end{equation}
For fixed $\Omega$ and $J_{0}$ (i.e.\ for a fixed value of the bulk $T_\mathrm{c}$), $\kappa^{-1}$ increases as $\sqrt{J_{1}}$. Because the difference between the polarizations at the film surface and interior depends on $\kappa^{-1}$,
the depolarizing field also grows with $J_1$; it then follows immediately that $P_\mathrm{ave}$ decreases as $J_1$ increases. This suppression is illustrated in \fref{fig:Pdep}(a), which shows the dependence of both the average polarization and $\kappa^{-1}$ on $J_1$. The polarization equals its bulk value when $J_1=0$ and drops as $J_1$ increases. Notably, there is a critical value of $J_1$ (which depends on the number of layers, $N_L$, in the film) above which ferroelectricity is completely suppressed. For the 50-layer film modelled here, this value is approximately 17~meV.

Alternatively, one can fix $J_1$ and consider how $P_\mathrm{ave}$ depends on film thickness, as shown in \fref{fig:Pdep}(b). Here polarization increases and asymptotically approaches the bulk value with increasing $N_L$.  Ferroelectricity is completely suppressed below a critical film thickness, with the value of this critical thickness depending on $J_1$. The results shown in \fref{fig:Pdep}(b) are for $J_1=10$~meV, and give a critical thickness of 30 layers. For $J_1=100$~meV, the critical thickness is closer to 300 layers.

Finally, the effect of increasing $J_{0}$ is shown in \fref{fig:Pdep}(c). Because the bulk value of polarization $P_\mathrm{bulk}$ depends on $J_0$, we show the ratio $P_\mathrm{ave}/P_\mathrm{bulk}$ as a function of $J_0/\Omega$. In bulk materials, the threshold for ferroelectricity is $J_0=$~$\Omega$, and this is increased by finite size effects in the 50-layer film as shown in \fref{fig:Pdep}(c). Size effects quickly become unimportant with increasing $J_0$, as $P_\mathrm{ave}$ rapidly increases towards its bulk value. Indeed, when $J_0$ is only twice $\Omega$, $P_\mathrm{ave}=0.93P_\mathrm{bulk}$.

These calculations show that doped quantum paraelectrics such as Sr$_{1-x}$Ca$_x$TiO$_3$, which have $J_0$ close to $\Omega$, should be highly sensitive to film thickness in the short-circuit geometry. While this might be naively anticipated based on the argument that the correlation length $\xi$ is comparable to the film thickness near a ferroelectric transition, this argument is wrong because the relevant length $\kappa^{-1}$ is actually rather short and does not diverge at the quantum critical point. Rather, the sensitivity is due to depolarizing fields, which can easily overwhelm the weak ferroelectricity.

\section{(001) LAO/STO Interface}
\label{sec:interface}

In the final section of this work, we apply the modified TIM to the (001) LAO/STO interface. For this calculation, the Hamiltonian must include an electronic term that describes the 2DEG that forms at the interface. The total Hamiltonian is thus
\begin{equation}
\hat{H} = \hat{H}_\mathrm{e} + \hat{H}_\mathrm{TIM},
\end{equation}
where $\hat{H}_\mathrm{TIM}$ is given by equation~\eref{TIM_full} and $\hat{H}_\mathrm{e}$ is the electronic term discussed below. These two terms are linked through the electric field, which appears explicitly in $\hat{H}_\mathrm{TIM}$, and appears implicitly in $\hat{H}_\mathrm{e}$ through the electrostatic potential.

We outline the calculations in \sref{sec:int_method}, and show results for the effect of $J_{1}$ on the interfacial 2DEG in \sref{sec:int_res}. The main result from this section is that the conventional and modified TIM make very different predictions for the structure of the 2DEG.

\subsection{Method} \label{sec:int_method}
We assume that the 2DEG arises due to a combination of top gating and the polar catastrophe. In this case a total charge density $-en_\mathrm{LAO}$ is donated from the LAO surface to the interface, where $n_\mathrm{LAO}$ is the surface hole density, in order to neutralize the polar discontinuity between the two materials. Top gating gives control over the number of free electrons doped into the system.

\begin{figure}[]
	\centering
	\includegraphics[width = 0.7\linewidth]{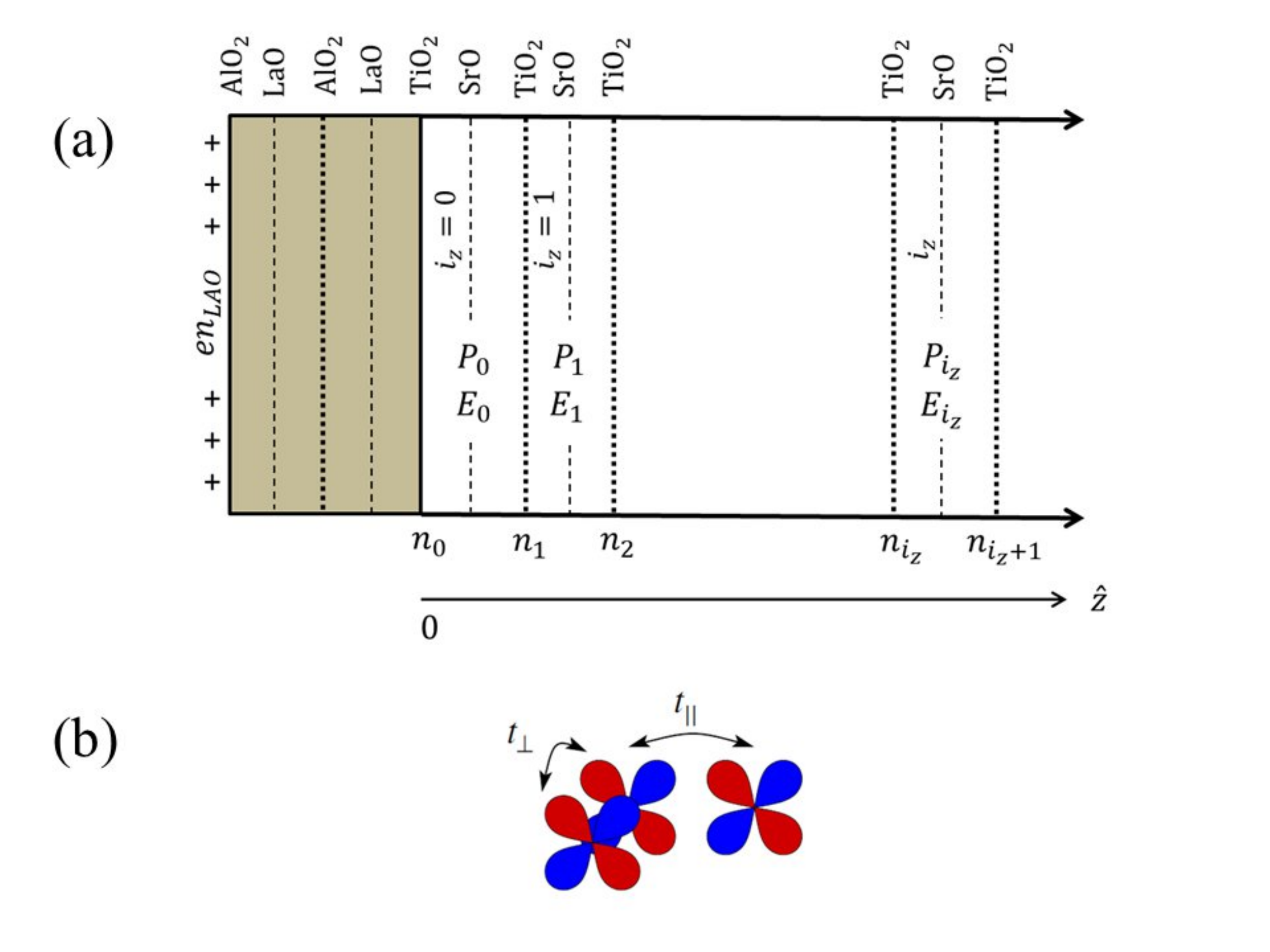}
	\caption{Structure of the LaAlO$_{3}$/SrTiO$_{3}$ interface. (a) The SrTiO$_3$ substrate is discretized, with each layer a single unit cell thick. The electron density $n_{i_{z}}$ is confined to the $i_z$th TiO$_2$ plane, which makes up the left face of layer $i_z$. The regions between the TiO$_2$ layers are treated as a polarizable medium that is modelled by the TIM. The positive charge density $+en_\mathrm{LAO}$ at the LAO surface is compensated by an equal but opposite charge in the two-dimensional electron gas (2DEG). (b) Electrons in the 2DEG hop between neighbouring Ti $t_{2\mathrm{g}}$ orbitals of the same type. The hopping amplitude between orbitals is either strong ($t^{\parallel}$) if the hopping path is in the plane of the orbitals, or weak ($t^{\perp}$) if the hopping path is perpendicular to the plane of the orbitals. Panel (b) is re-published from \cite{atkinson17}.}
	\label{fig:LAOSTO}
\end{figure}

As shown in \fref{fig:LAOSTO}(a), we adopt a discretized model comprising alternating metallic TiO$_2$ layers with electron densities $n_{i_z}$ and dielectric layers with polarizations $P_{i_z}$.  Translational invariance is assumed within the $xy$-plane, but not along the $z$-axis perpendicular to the interface. The system's properties are therefore only dependent on layer.

The 2DEG is composed of electrons that occupy  titanium $t_{2\mathrm{g}}$ orbitals in the STO substrate. Although the unit cell is tetragonally distorted both by unit cell rotations about the $c$-axis and by interfacial strains, to a good approximation we can assume STO has the cubic structure typical of a perovskite material, as shown in the inset of \fref{fig:STOX_compspin}. We adopt a tight-binding model in which the conduction bands are made up of $t_{2\mathrm{g}}$ orbitals \cite{raslan17,Stengel:2011hy,Khalsa:2012fu}, and assume that electrons only hop between orbitals of the same type (ie. from one $d_{xz}$ orbital to another $d_{xz}$ orbital; other hopping matrix elements vanish in the cubic phase by symmetry, and are generally small when lattice distortions are included).

\subsubsection{Electronic Hamiltonian}

The electronic Hamiltonian is made up of a hopping kinetic energy $\hat{T}$ and an electrostatic potential energy $\hat{U}$:
\begin{equation}
\hat{H}_\mathrm{e} = \hat{T} + \hat{U}.
\end{equation}
The hopping energy is 
\begin{eqnarray} \label{T_k}
\hat{T} = \sum_{i_{z} \bt{k} \alpha \sigma} \epsilon_{i_{z} \alpha} \hat{c}^{\dag}_{i_{z} \bt{k} \alpha \sigma} \hat{c}_{i_{z} \bt{k} \alpha \sigma} + \sum_{\langle i_{z},i_{z}' \rangle \alpha \sigma} \sum_{\bt{k} \bo{\delta}} t^{\alpha}_{\bo{\delta}} e^{-i \bt{k} \cdot \bo{\delta}} \hat{c}^{\dag}_{i_{z}' \bt{k} \alpha \sigma} \hat{c}_{i_{z} \bt{k} \alpha \sigma},
\end{eqnarray}
where $\hat{c}^{\dag}_{i_{z} \bt{k} \alpha \sigma}$ creates an electron with spin $\sigma$ and orbital type $\alpha$ in the 2D plane-wave state $\bt{k}=(k_{x}, k_{y})$ in layer $i_{z}$. $\sum_{\langle i_{z}, i_{z}' \rangle}$ is a sum over nearest-neighbour layers $i_{z}$ and $i_{z}'$. $t^{\alpha}_{\bo{\delta}}$ is the hopping matrix element for an electron in orbital type $\alpha$ hopping along path $\bo{\delta}$ to a nearest-neighbour site. $\epsilon_{i_{z} \alpha}$ is the atomic energy of an orbital site in layer $i_{z}$, and can be set to zero in calculations.

In the tight-binding model, there are six possible hopping paths. Hopping along $\hat{x}$ corresponds to a displacement $\bo{\delta}_{x}=(\pm a, 0, 0)$ and hopping amplitude $t^{\alpha}_{x}$, and so on for hopping along $\hat{y}$ and $\hat{z}$. Then, equation~\eref{T_k} simplifies to
\begin{eqnarray} \label{T_full}
\fl \hat{T} = \sum_{i_{z} \bt{k} \alpha \sigma} \Big( \epsilon_{\bt{k} \alpha} \hat{c}^{\dag}_{i_{z}\bt{k}\alpha \sigma} \hat{c}_{i_{z}\bt{k}\alpha \sigma} + t^{\alpha}_{z} \hat{c}^{\dag}_{i_{z}+1, \bt{k}\alpha \sigma} \hat{c}_{i_{z}\bt{k}\alpha \sigma} + t^{\alpha}_{z} \hat{c}^{\dag}_{i_{z}-1, \bt{k}\alpha \sigma} \hat{c}_{i_{z}\bt{k}\alpha \sigma} \Big),
\end{eqnarray}
where $\epsilon_{\bt{k} \alpha} = 2 t^{\alpha}_{x} \cos (k_{x} a) + 2 t^{\alpha}_{y} \cos (k_{y} a)$. As illustrated in \fref{fig:LAOSTO}(b), the amplitudes $t^{\alpha}_{x}, t^{\alpha}_{y}$ and $t^{\alpha}_{z}$ are denoted by $t^\|$ for hopping paths that lie in the plane defined by $\alpha$, and $t^\perp$ for hopping paths that are perpendicular to this plane. We take $t^{\parallel}=-0.236$~eV and $t^{\perp}=-0.035$~eV as in \cite{raslan17}.

The electrostatic potential energy is due to the charge on the LAO surface, the 2DEG, and the bound charge due to the polarization of the STO:
\begin{equation} \label{U_k}
\hat{U} = -e \sum_{i_{z} \bt{k} \alpha \sigma} V_{i_{z}} \hat{c}^{\dag}_{i_{z}\bt{k}\alpha \sigma} \hat{c}_{i_{z}\bt{k}\alpha \sigma},
\end{equation}
where $e$ is electron charge and $V_{i_{z}}$ is the electrostatic potential in layer $i_{z}$.

Combining equations~\eref{T_full} and \eref{U_k} gives the full electronic Hamiltonian:
\begin{eqnarray} \label{Hel_full}
\hat{H}_\mathrm{e} &=& \sum_{i_{z} \bt{k} \alpha \sigma} \Big\lbrace \Big( \epsilon_{\bt{k} \alpha} - e V_{i_{z}}  \Big) \hat{c}^{\dag}_{i_{z}\bt{k}\alpha \sigma} \hat{c}_{i_{z}\bt{k}\alpha \sigma} \nonumber \\ && + t^{\alpha}_{z} \hat{c}^{\dag}_{i_{z}+1, \bt{k}\alpha \sigma} \hat{c}_{i_{z}\bt{k}\alpha \sigma} + t^{\alpha}_{z} \hat{c}^{\dag}_{i_{z}-1, \bt{k}\alpha \sigma} \hat{c}_{i_{z}\bt{k}\alpha \sigma} \Big\rbrace.
\end{eqnarray}
The Hamiltonian can be written as an $N_L\times N_L$ matrix in the layer index, $\hat{H}_\mathrm{e} =$~$\sum_{\bt{k} \alpha \sigma} \bo{\hat{c}}^{\dag}_{\bt{k} \alpha \sigma} \bt{H}_{\bt{k} \alpha \sigma} \bo{\hat{c}}_{\bt{k} \alpha \sigma}$, with $\bo{\hat{c}}_{\bt{k} \alpha \sigma}  =$~$(\hat c_{0\bt{k} \alpha \sigma}, \ldots, \hat c_{N_L-1, \bt{k}\alpha \sigma})$ and 
\begin{equation}
\bt{H}_{\bt{k} \alpha \sigma} = \bt{H}_\alpha + \epsilon_{\bt{k} \alpha} \bt{I},
\end{equation}
where $\bt{H}_\alpha$ is independent of $\bt{k}$ and $\bt{I}$ is the identity matrix. The eigenergies are particularly simple, with
\begin{equation}
\epsilon_{n\bt{k}\alpha} =  \lambda_{n\alpha} + \epsilon_{\bt{k} \alpha}, 
\end{equation}
where $\lambda_{n\alpha}$ are the eigenvalues of $\bt{H}_\alpha$ and  $n$ is the band index. The eigenvectors of $\bt{H}_{\bt{k} \alpha \sigma} $, which represent the layer-dependent wavefunctions, are $\bt{k}$-independent and satisfy
\begin{equation}
\sum_{j_z} [\bt{H}_\alpha]_{i_z j_z} \psi_{j_z n \alpha} = \lambda_{n\alpha}  \psi_{i_z n \alpha}. 
\end{equation}

From this, the free electron density (per unit cell) in layer $i_{z}$ is
\begin{equation} \label{n_i}
n_{i_{z}} = \frac{1}{ N} \sum_{n \bt{k} \alpha \sigma} f_\mathrm{FD} (\epsilon_{n \bt{k} \alpha}) | \psi_{i_{z} n \alpha} |^{2},
\end{equation}
where $N$ is the total number of $k_{x}$- and $k_{y}$-points, and $f_\mathrm{FD} (\epsilon_{n \bt{k} \alpha})$ is the Fermi-Dirac distribution.

We note that the mean-field equations described in this section neglect thermal fluctuations of both the lattice and the charge density. Both of these broaden the electronic spectral functions, as in Fermi liquid theory, and can in principle mix the bands. These are perturbative effects, however, and band structure calculations like the one outlined here generally provide a good quantitative description of the electronic structure, even at room temperature.

\subsubsection{Electric Field}

The electric potential in layer $i_{z}$ is obtained by integrating the electric field from layer 0 to layer $i_z$, which sets the interface to be the zero of potential. Then,
\begin{equation} \label{V}
V_{i_{z}} = -a \sum_{j_{z} < i_{z}} E_{j_{z}} + V_{0},
\end{equation}
with $a=3.902$~\AA{} the STO lattice constant.

Just as in \sref{sec:FEfilm}, the electric field can be obtained using Gauss' law, 
\begin{equation} \label{Gauss_int}
\frac{d}{dz} (\epsilon_{\infty} E(z) + P(z)) = \rho^\mathrm{2DEG}(z) + \rho^\mathrm{ext}(z),
\end{equation}
where $\rho^\mathrm{2DEG}(z)$ is the free charge density and $\rho^\mathrm{ext}$ is the external charge density along the LAO surface. The polarization $P(z)$ is obtained from the modified TIM.

Within the discretized model, the electrons are treated as if they are confined to two-dimensional TiO$_2$ layers, so
\begin{equation}
\rho^\mathrm{2DEG}(z) = -\frac{e}{a^{2}} \sum_{i_{z}} n_{i_{z}} \delta(z - i_z a),
\end{equation}
where $n_{i_{z}}$ is given by equation~\eref{n_i}. Similarly, the external charge density is confined to the top LAO layer,
\begin{equation}
\rho^\mathrm{ext} = \frac{en_\mathrm{LAO}}{a^{2}} \delta(z - z_\mathrm{LAO}),
\end{equation}
where $z_\mathrm{LAO}$ is the distance from the interface to the LAO surface. Now, integrating equation~\eref{Gauss_int} over $z$ gives the electric field in layer $i_{z}$:
\begin{equation} \label{E_int}
\epsilon_{\infty} E_{i_{z}} = - P_{i_{z}} - \frac{e}{a^{2}} \sum_{j_{z} \leq i_{z}} n_{j_{z}} + \frac{en_\mathrm{LAO}}{a^{2}},
\end{equation}
which is required for the TIM [equation~\eref{TIM_full}] and the electric potential [equation~\eref{V}].

\subsection{Results} \label{sec:int_res}

Here, we explore the effect that $J_{1}$ has on the electron distribution, eigenenergies, polarization and potential energy for the (001) LAO/STO interface. As a key point of comparison, these calculations include the case $J_1=0.65$~meV ($J_\mathrm{an}=0$), which corresponds to the conventional TIM, in order to clearly highlight why the modified TIM requires the term introduced in equation~\eref{TIM_full} to correctly model interfaces.

\begin{figure}[tb]
	\centering
	\includegraphics[width=0.7\linewidth]{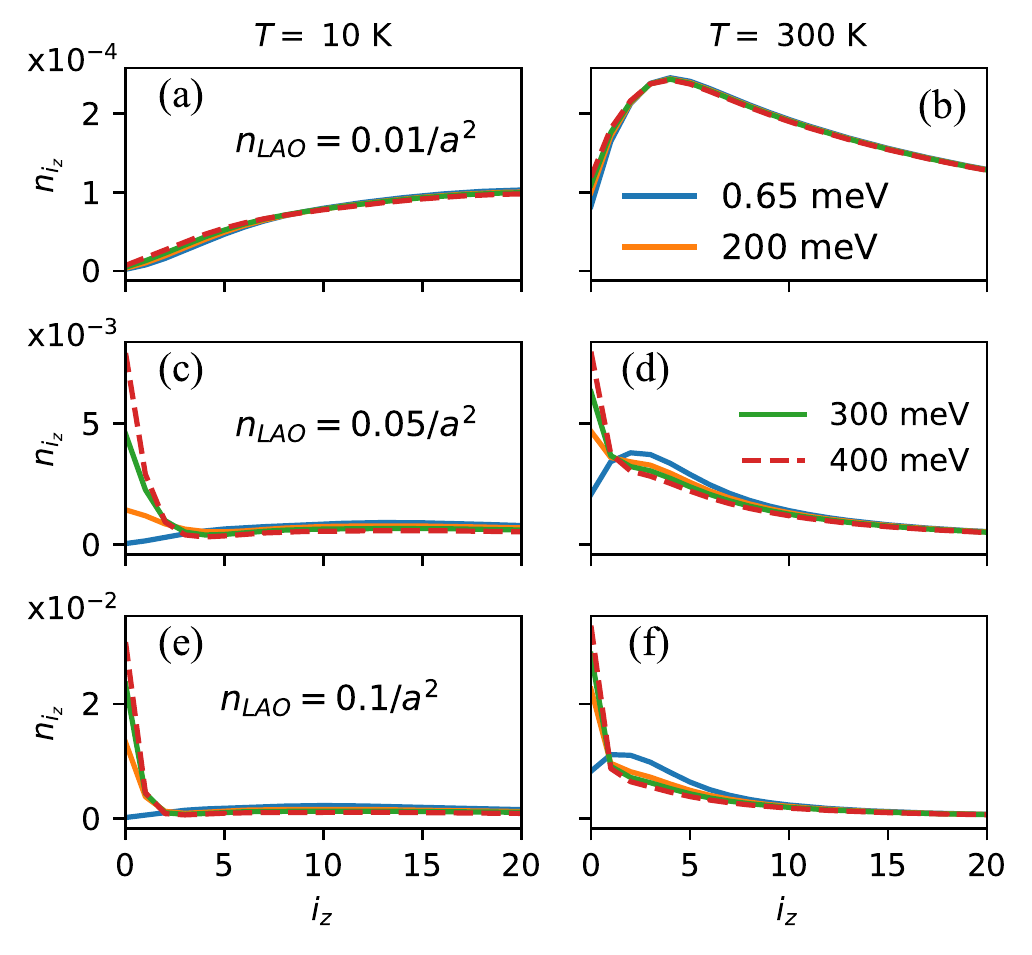}
	\caption{Electron density (per unit cell) for the first 20 layers of a 200-layer film. Results are for different $J_{1}$ values, at $T=10$~K and 300~K, and (a)-(b) $n_\mathrm{LAO}=0.01/a^{2}$, (c)-(d) $n_\mathrm{LAO}=0.05/a^{2}$, and (e)-(f) $n_\mathrm{LAO}=0.1/a^{2}$.}
	\label{fig:int_nvsN}
\end{figure}

Previous work has established that the 2DEG is composed of both interfacial and tail components. The interfacial component is tightly confined to the interface, and appears as a peak in the electron density extending over the first few layers of the substrate, while the tail component  extends far into the STO substrate \cite{copie09,dubroka10,park13,gariglio15,raslan18}. Except at the very lowest dopings, the majority of the electrons are confined close to the interface, with as many as 70\% of the electrons in the 2DEG found in approximately the first 10~nm \cite{raslan17,copie09}. This interfacial peak in the electron density is strongly temperature- and electron doping-dependent, with the electrons spreading further out into the STO as temperature or doping decreases \cite{raslan17}. The first $d_{xy}$ band contributes the most electrons to the interface states, while the first $d_{yz}$ and $d_{xz}$ bands make up the majority of the tail states and are seen to have the most temperature-dependence \cite{raslan17}.

\begin{figure}[]
	\centering
	\includegraphics[width=0.7\linewidth]{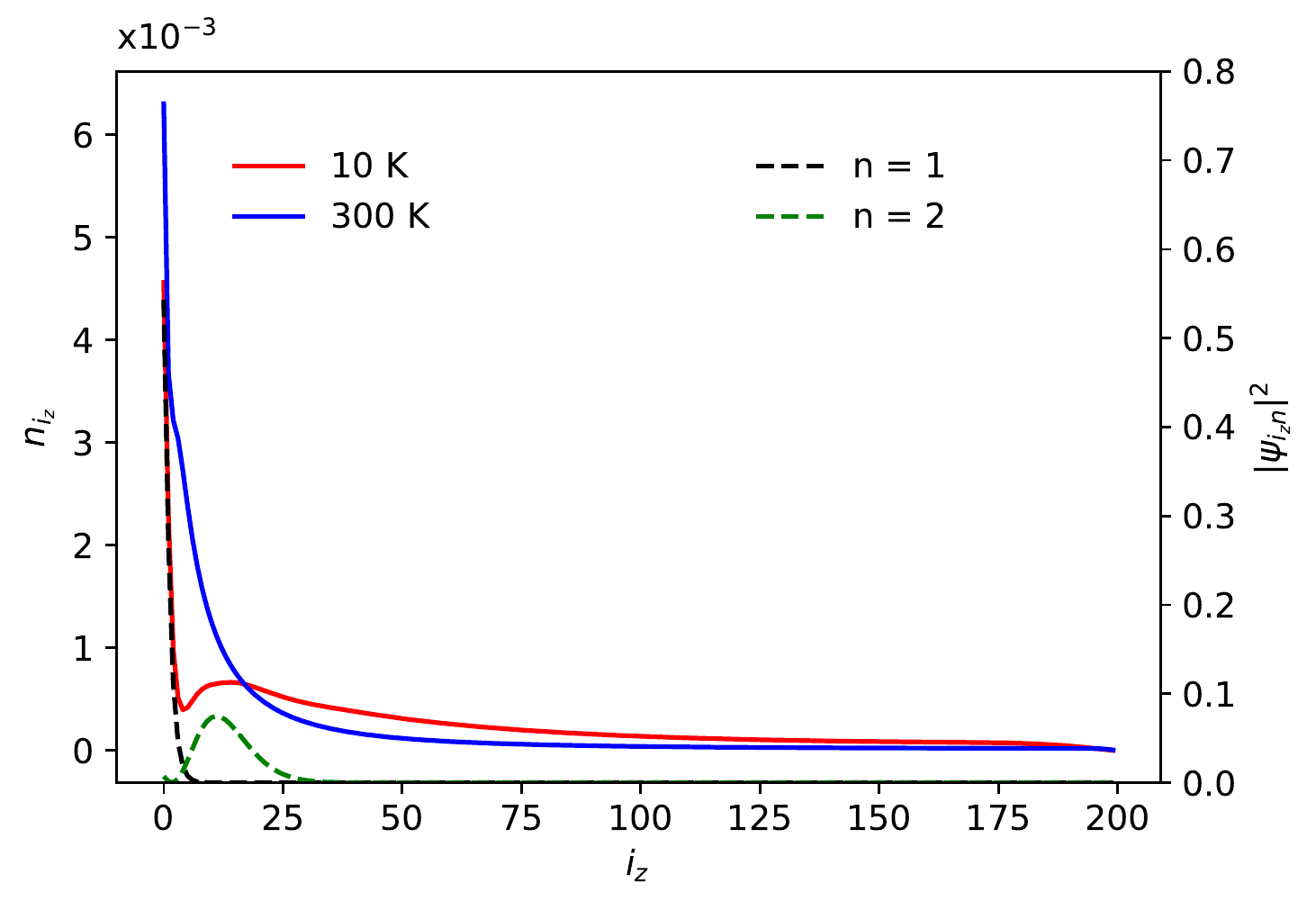}
	\caption{Profile of the electron density (solid) for $n_\mathrm{LAO}=0.05/a^{2}$ and $J_{1}=300$~meV, at both 10~K and 300~K. Probability distributions (dashed) for the two lowest-energy $d_{xy}$ bands ($n = 1,2$) are shown for 10~K.}
	\label{fig:int_nvsN_one}
\end{figure}

The electron density is plotted in figures~\ref{fig:int_nvsN} and \ref{fig:int_nvsN_one}. \Fref{fig:int_nvsN} explores the effect $J_{1}$ has on the electron density, focusing particularly on the interface region, while \fref{fig:int_nvsN_one} shows the full profile over the entire film for a typical set of model parameters.

We begin analyzing \fref{fig:int_nvsN} by focusing on the results of the conventional TIM. When $J_1=0.65$~meV, there is no evidence that electrons are confined to the interface region at 10~K at any doping, in disagreement with experiments. Weak confinement does appear at 300~K due to the reduced dielectric susceptibility at high $T$, and the 2DEG does move towards the interface with increasing doping; however, the density is expected to be strongly peaked at the interface, and this is not seen. The conventional TIM, therefore,  does not capture the physics of STO interfaces.

The remaining curves in \fref{fig:int_nvsN} show how the charge profile changes with increasing $J_1$.  These results are for fixed $J_0$ and $\Omega$ (which determine the uniform dielectric susceptibility), and the only difference between the curves is therefore the correlation length $\xi$. These curves show that increasing $J_{1}$ (or equivalently, increasing $\xi$) tends to increase electron density at the interface, except at the lowest doping.

At the lowest doping,  $n_\mathrm{LAO}=0.01/a^{2}$, $J_{1}$ has little effect on the electron density at both high and low temperature. Indeed, interface states are absent  for all $J_1$ values up to 400~meV. While this lack of interface states is consistent with previous calculations \cite{raslan18}, it is not consistent with experiments \cite{yin19,joshua12}, and likely points to some additional missing physics in the model \cite{raslan18}.

At intermediate doping, $n_\mathrm{LAO}=0.05/a^{2}$, the electron density does develop an interfacial component as $J_{1}$ increases. This interfacial state extends only a few unit cells from the interface, and is more tightly confined at large $J_1$. There is thus a clear qualitative distinction between the modified and conventional TIMs in this case. At high doping, $n_\mathrm{LAO} = 0.1/a^{2}$, the trends are similar. The electron density is confined closer to the interface and is less strongly temperature dependent than at lower doping, at least when $J_1\geq200$~meV. Both of these trends are consistent with results reported in \cite{raslan17}. 

\begin{figure}[]
	\centering
	\includegraphics[width=0.7\linewidth]{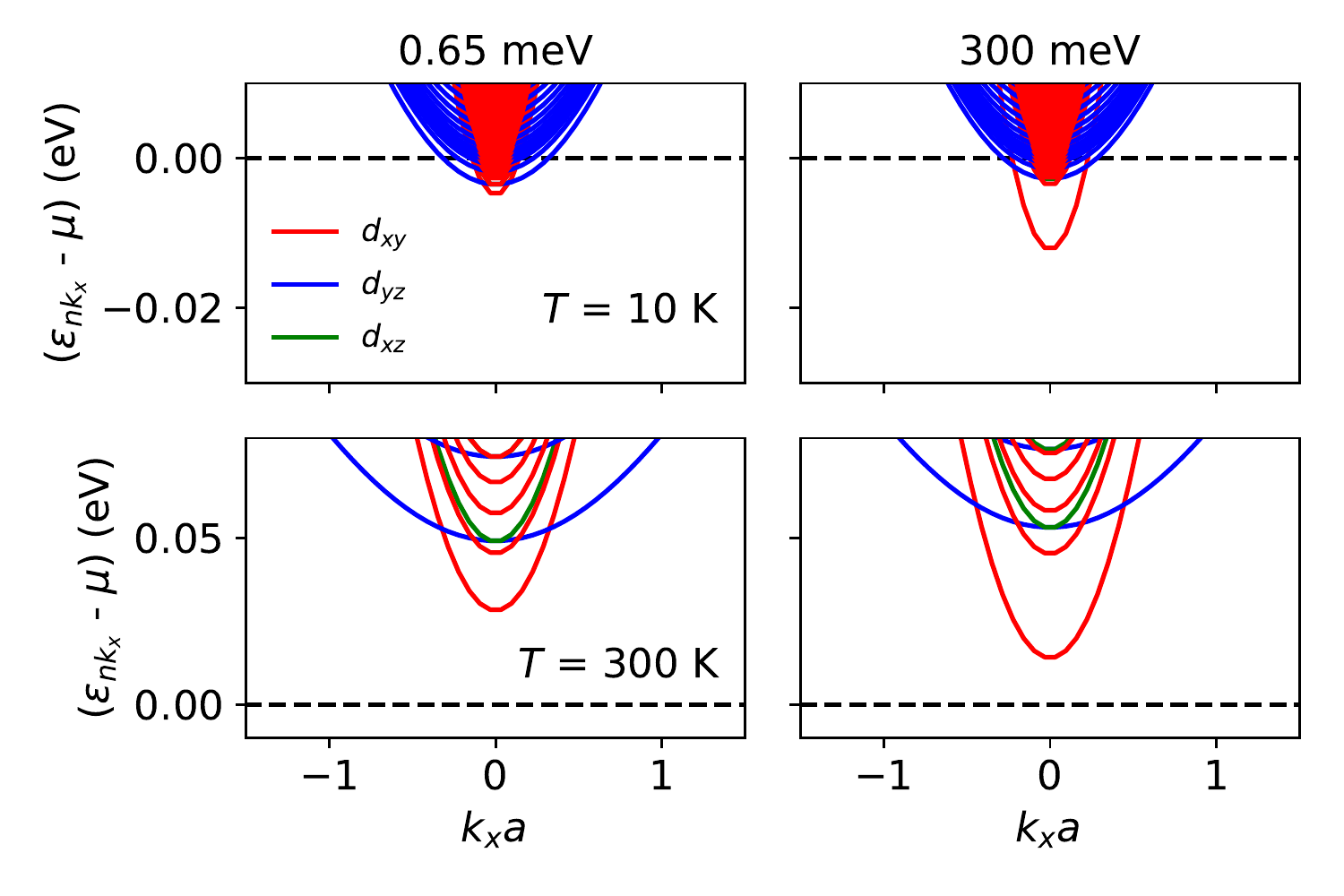}
	\caption{Band structure for $n_\mathrm{LAO}=0.05/a^{2}$ at $J_1=0.65$~meV and 300~meV. Results are shown for a 200-layer substrate at both 10~K (top) and 300~K (bottom).}
	\label{fig:int_bands05}
\end{figure}

\Fref{fig:int_nvsN_one} shows the electron density across the full thickness of the STO film for a typical $J_1$ value at intermediate doping for both 10~K and 300~K. We choose the value of $J_{1}=300$~meV as physically reasonable based on the results in \fref{fig:int_nvsN}. At 10~K, the charge profile shows a peak-dip-hump structure that has not been reported in previous calculations. To understand its origin, we plot also the wavefunctions $|\psi_{i_z n \alpha}|^2$ for the first ($n=1$) and second ($n=2$) $d_{xy}$ bands ($\alpha=xy$) at 10 K. These show that the dip comes from the extremely tight confinement of the first $d_{xy}$ band to the interface.

The band structure is shown in \fref{fig:int_bands05} for intermediate doping for both the conventional TIM and the modified TIM ($J_1= 300$~meV). At 10~K, the band structures of the two models are  quasi-continuous, which is indicative of deconfined tail states, except for a single $d_{xy}$ band that sits below the continuum in the modified TIM, and which corresponds to the interface state discussed above. At high $T$, the band structures are discrete, which is indicative of confinement to the interface region. At this temperature, the effects of $J_1$ are quantitative, rather than qualitative.

\begin{figure}[tb]
	\includegraphics[width=0.5\linewidth]{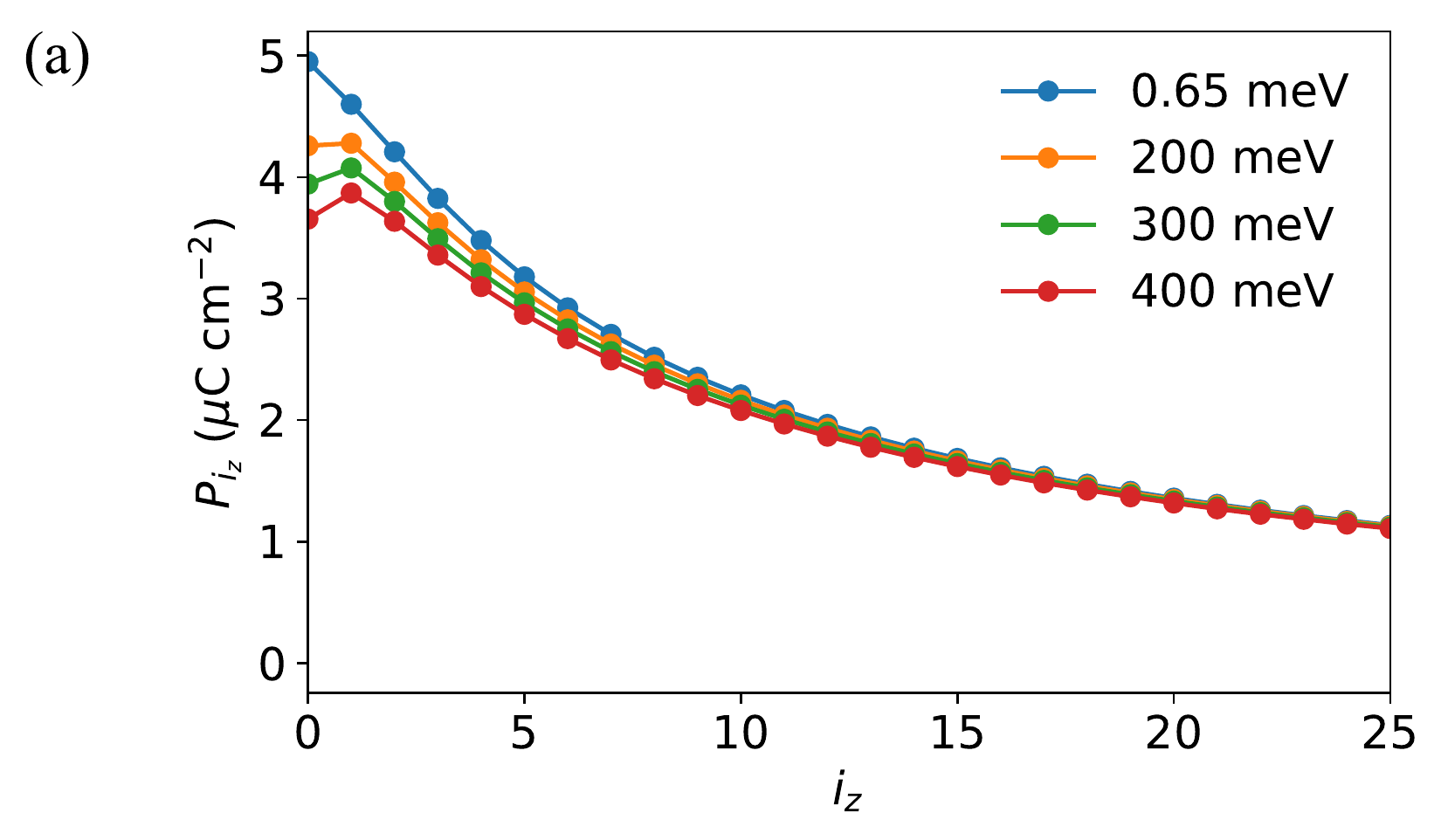}
	\includegraphics[width=0.5\linewidth]{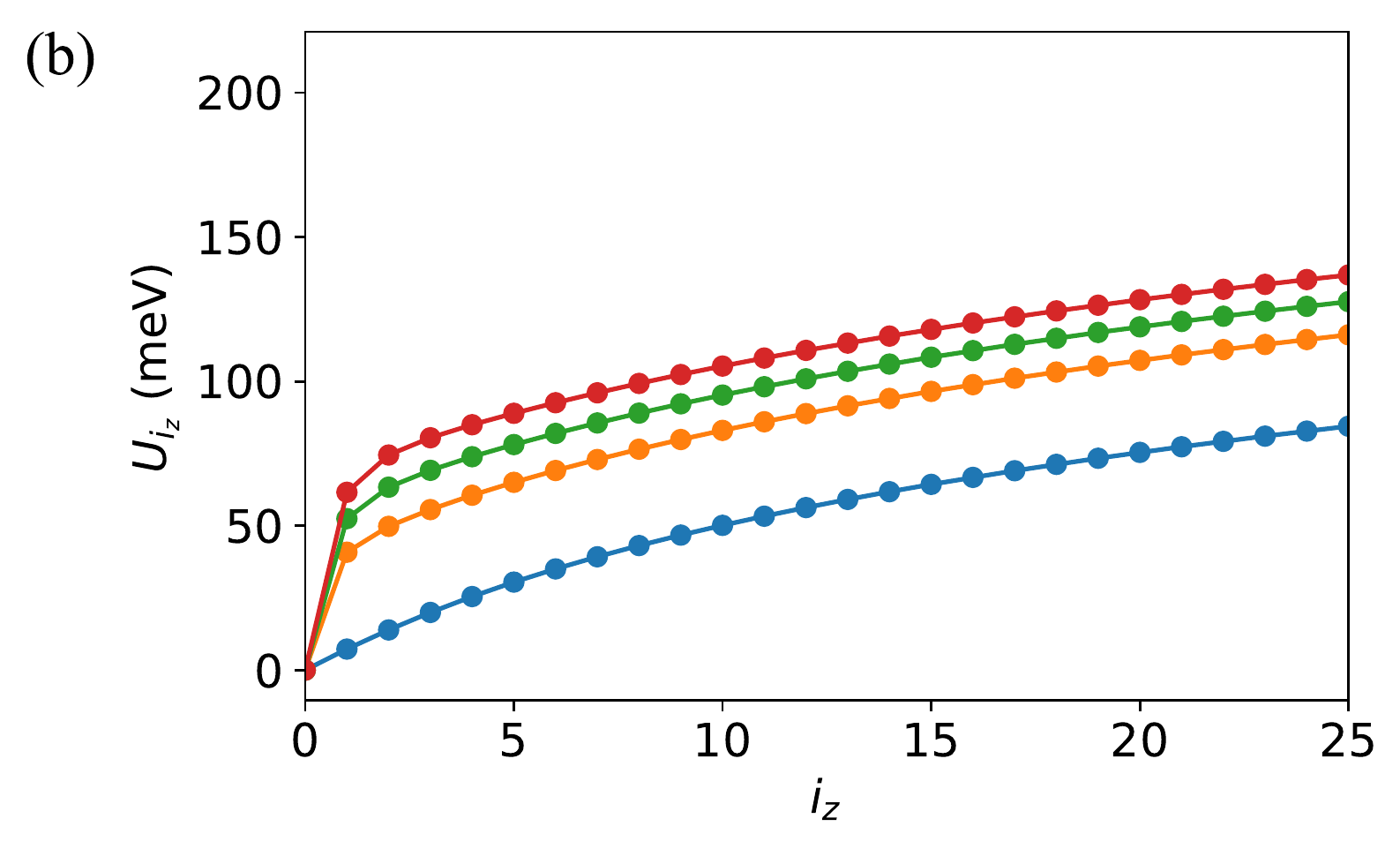}
	\caption{(a) Polarization and (b) electron potential energy in the first 25 layers of a 200-layer SrTiO$_3$ substrate for different $J_{1}$ values at 300~K and $n_\mathrm{LAO}=0.05/a^{2}$.}
	\label{fig:int_PvsN}
\end{figure}

Finally, we plot in \fref{fig:int_PvsN} the polarization and potential energy at 300~K for intermediate doping. These plots show that there are clear distinctions between the conventional and modified TIMs in the interfacial region. In particular, the polarization near the interface is reduced, by up to 25\%, as $J_1$ increases. This reduction is similar to that discussed in the case of the thin film, with one key difference: because electric fields are screened by the 2DEG, the relevant length scale over which differences between the curves decay in \fref{fig:int_PvsN}(a) is $\xi$, and not $\kappa^{-1}$ \cite{atkinson17}.

Similar to the ferroelectric thin films discussed in \sref{sec:FEfilm}, this reduced polarization incompletely screens the electric fields produced by the LAO surface charge and results in a large  field at the interface. This is reflected in the potential energy profiles shown in \fref{fig:int_PvsN}(b). In particular, large values of $J_1$ generate a deep potential well that confines the lowest $d_{xy}$ band tightly to the interface. On the other hand, $J_{1}$ has little effect on the electric field away from the interface, and so each potential energy curve has  roughly the same slope for $i_{z} > 2$. In summary, \fref{fig:int_PvsN} illustrates the mechanism by which the anisotropic pseudospin term in the modified TIM generates the interfacial component of the 2DEG that is observed at LAO/STO interfaces.

\section{Conclusions}

We showed that the conventional transverse Ising model misses key features of spatially inhomogeneous STO-based nanostructures. To fix this we modified the TIM by adding an anisotropic pseudospin energy to the Hamiltonian. This corrects a deficiency of the TIM, namely that if one fits the model parameters to the bulk (homogeneous) susceptibility, the polarization correlation length is also fixed by the model and is at least an order of magnitude smaller than it should be.  

To illustrate the effects of the new term, we considered two applications of the modified TIM: first, to thin films of an STO-like ferroelectric; and second, to a metallic LAO/STO interface. In both cases, the key point is that the conventional TIM underestimates the reduction of the polarization due to the surface; this reduced polarization leads to a reduced screening of electric fields in the interface region, which in turn has profound effects on the film or interface. In the case of the ferroelectric film, these fields depolarize the polarization in the film; in the case of the interface, they create a confining potential that generates tightly bound interface states.

\ack
This work has been supported by the Natural Sciences and Engineering Research Council (NSERC) of Canada. 

\section*{References}
\bibliographystyle{iopart-num}
\bibliography{refs}

\end{document}